# New Strategies for Solar Cells Beyond the Visible Spectral Range

*Fabio Marangi [1,2]\*, Matteo Lombardo [1], Andrea Villa [1], Francesco Scotognella [1,2]*

1 Department of Physics, Politecnico di Milano, Piazza L. da Vinci 32, 20133 Milano, Italy

2 Center for Nano Science and Technology@PoliMi, Istituto Italiano di Tecnologia (IIT), Via Giovanni Pascoli, 70/3, 20133, Milan, Italy

\* Corresponding Author: fabio.marangi@polimi.it

## Abstract

The endeavor of the scientific community to maximize the possibility to harvest Sun irradiation for energy production is mainly devoted to the improvement of the power conversion efficiency of devices and to the extension of the spectral range in which solar devices operate. Considering that a significant portion of the Sun irradiation at the ground level is in the infrared, the research on materials and systems that operate in such region is gaining increasing attention. In this review, we will report recent advancements in multijunction solar cells, inorganic-organic perovskite solar cells, organic solar cells, colloidal quantum dot solar cells focusing on the absorption of such devices in the infrared. In addition, the use of upconverting nanostructures will be introduced as a way to indirectly exploit infrared radiation to increase power conversion efficiency of photovoltaic devices. Moreover, we will describe plasmon induced hot electron extraction based solar cells, that are particularly promising in absorbing the infrared portion of the Sun irradiation when the active materials are doped semiconductors, which show intense plasmonic resonances in the infrared. The review includes the optical spectroscopy tools to study the hot electron extraction from doped semiconductor-based heterojunctions.

**Keywords:** Hot electrons, infrared photovoltaics, plasmonics, spectroscopy

## 1. Introduction

The energy crisis the whole world must face nowadays induced many scientists to focus their research on the use of sources that have been neglected for ages and on increasing the efficiency of energy conversion of already existing technologies in order to make it up for the demanding needs of our society, which are continuously growing. According to Nikolaidou et al. [1] a conservative estimate predicts a 35% increase in energy consumption in the next two decades, raising our power needs to an appalling 30 TW in 2050. Moreover, the studies in the field of renewable energy will play a crucial role in providing electricity to places that have limited access to continuous power supplies such as those located in remote regions or those with limited resources. The research in this field will also assist in reducing $CO_2$ emissions and fight global warming and polluting by increasing the use of greener technologies to power electronic devices, industrial equipment and vehicles intended for humans or goods transportation [2]. For an extensive review on the environmental and energetic problem refer to [3].

In terms of economic resources, among those that have been defined renewable energy sources (i.e. biomass and waste, biofuels, geothermal, solar, hydro and wind) the majority of investments have been made towards the transitioning to the generation of electric power through wind and solar energy. This is also due to the fact that wind and solar energy are globally available and are not subjected to geography as it is instead for geothermal and hydropower. In fact, even if hydroelectric energy has very high conversion efficiency, its use is limited due to the issue of constructing plants close to water resources and in places with given geographical characteristics. On the contrary solar panels are easily adaptable to any environment, including the urban one where they can be placed on roofs or on surfaces constantly exposed to sunlight.

In modern buildings, windows play an important role not only as an architectural feature, but also in terms of energy consumption. In fact, clever positioning of windows in buildings is crucial to reduce energy demands in terms of heating, cooling and of course lighting. [4] In this framework windows could also be integrated with transparent or semi-transparent photovoltaics and contribute to electricity production in order to have perfectly optimized buildings in terms of energy saving. [5]

Efficient photovoltaics beyond the visible spectral range will also be a breakthrough for solar technologies. In fact, thanks to their transparency in such range, those devices will be easily integrated with silicon-based (or any other visible-absorbing material) solar panels, solving the problem of spectral losses and without interfering in any way with silicon absorption. [6] In particular, plasmonic solar cells based on doped semiconductor nanocrystals seem to be a valid alternative to conventional photovoltaics, focusing on the absorption of the Near-Infrared (NIR) part of the solar spectrum. [7] Those devices rely on the generation of hot electrons in the plasmonic material, which can be subsequently injected in the conduction band of a neighboring semiconductor and in the end extracted from an electrode.

An alternative strategy to achieve light conversion in the range between 0.7 eV and 0.4 eV is the employment of thermophotovoltaic systems and solar thermoelectric generators. In the case of thermophotovoltaic systems, low bandgap semiconductors can harvest thermal energy and potential candidates are GaSb, Ge, InGaAs, and InGaAsSb. [8] In the case of solar thermoelectric generators, such solid-state heat engines convert sunlight directly into electricity through the thermoelectric effect or Seebeck effect and consist of a solar absorber and the thermoelectric generator. A typical thermoelectric generator is made of many thermocouples, where a p-type leg and an n-type leg are electrically connected in series and thermally in parallel. By applying a temperature gradient across the device, a voltage proportional to the number of thermocouples is generated, and a current stably sustained over a load, with a maximum power generation obtained at resistance matching. [9,10]

This review will mainly focus on reporting the efforts made for the fabrication of NIR-absorbing plasmonic photovoltaic devices. Different, non-plasmonic technologies implemented for the extension of the absorption of photovoltaic devices beyond or below the visible range will also be reported for the sake of completeness.

Chapter 2 reports the efforts for the fabrication of devices to harvest infrared radiation. The reasons behind such research will be explained and different materials such as III-IV-V groups semiconductors, organic-inorganic perovskite materials, organic dyes and semiconducting polymers and quantum-dots will be introduced. Those materials are commonly implemented for the fabrication of new generation solar cells and research in those fields now focuses on increasing efficiencies of the devices also by using infrared radiation. Followingly, upconverting nanoparticles will be briefly discussed. In this last case, infrared radiation is not directly used for charge generation, but it is absorbed by rare-earths elements nanostructures and re-emitted in the form of visible radiation to be exploited through visible light absorbing active layers.

In chapter 3, plasmon-induced hot electron extraction will be introduced as a mean to harvest infrared radiation. Basic optical properties of plasmonic nanostructures will be discussed in detail and different approaches to exploit such properties and implement plasmonic solar cells will be suggested.

Chapter 4 will briefly introduce some already existing plasmonic-semiconductor hybrid systems, while chapter 5 will report several techniques which can be useful to deeply understand the properties of plasmonic systems, in order to be able to exploit them for the fabrication of photovoltaic plasmonic devices with infrared-absorbing active layers.

## 2. Harvesting the IR Spectral Range

The sun is a huge source of energy. However, only part of it is actively exploited by photovoltaic devices to produce electric power. As a matter of fact, nearly 45% of solar radiation (Near-Infrared) (Figure 1), which at the ground level reaches around 1 kW/m$^2$, is lost due to spectral losses of silicon-based solar cells or problems regarding conversion efficiency. It is also in order to find a solution to this problem that the scientific community has put a lot of effort in the past two decades in the research revolving around photovoltaics.

Since the introduction into the market of single junction Si-based solar cells in 1950s (1$^{st}$ generation), researchers focused on finding a way to improve power conversion efficiency, both by investigating alternative device architectures and by making use of novel materials, also to reduce environmental impact of production processes, the amount of industrial waste and obtain state of the art, efficient devices. However single-junction solar cells are characterized by the detailed balance limit, or Shockley-Queisser limit. [11–13]

In fact, power conversion efficiency of single junction solar cells is thermodynamically limited to around 33,7% (AM 1.5 solar spectrum), depending on the concentration of incoming sunlight. AM 1.5 solar spectrum corresponds to a standardized condition for testing, taking into consideration that light reaching the surface of the solar panel goes through the atmosphere and gets attenuated and scattered (Figure 1). In particular, the limit arises mainly due to spectral losses which are intrinsic of a single junction device. Photons with energy below the bandgap of the material are not absorbed (red losses), while those with energy higher than the bandgap lose their excess energy as heat (blue losses). Being the sun a polychromatic light source, fixing the bandgap results in a tradeoff between the two losses. [13–15]

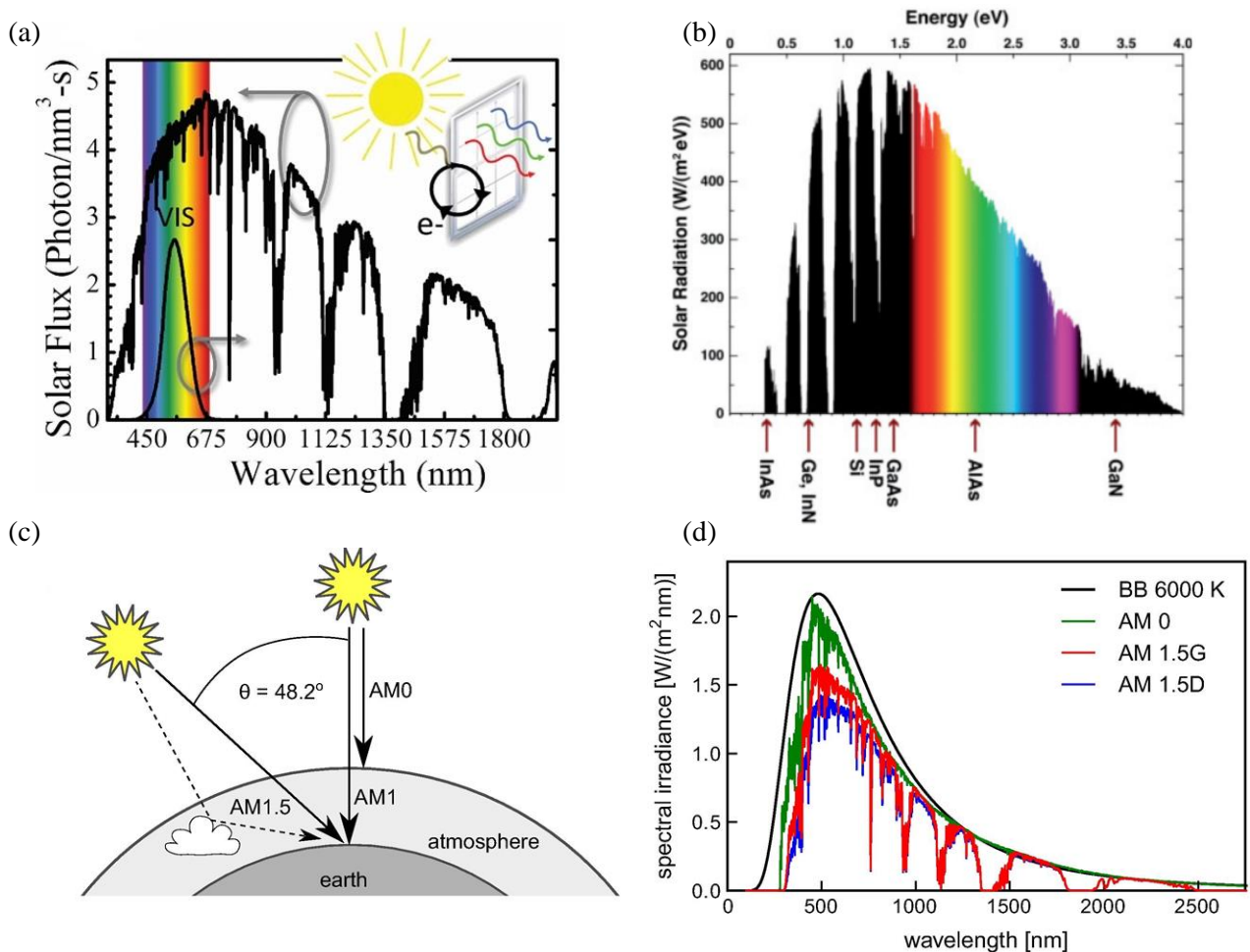

*Figure 1 - (a) Solar photon flux plotted versus wavelength. Only the visible part and a little part of the infrared can be harvested with Si-based solar cells. The inset shows a schematic of a visibly transparent window integrated with transparent photovoltaics. Adapted from [5] (b) The spectrum represents the intensity and spectral distribution of incoming sunlight versus energy (eV) for a given location and atmospheric conditions. The bandgaps of selected materials usually exploited for photovoltaics is also displayed. Adapted from [14] (c) Schematic representation of the spectral irradiance of the sun outside the earth's atmosphere (AM 0) and on the earth's surface for direct sunlight (AM 1.5 D - solid line) and direct sunlight with scattered contribution from the atmosphere (AM 1.5 G – solid and dashed line) integrated over a hemisphere. Adapted from [13] (d) Spectral irradiance according to ASTM G173-03 in comparison to the spectrum of a black body with surface temperature of 6000K used by Shockley and Queisser. Adapted from [13]*

## 2.1 Multijunction Solar Cell

One of the solutions implemented to overcome this thermodynamic limit is represented by multijunction solar cells. These devices are able to achieve higher conversion efficiencies (>33.7%) by separating the absorption of the polychromatic solar spectrum into semiconductors with different bandgaps, so that high energy photons are absorbed by a high band gap material (UV), while lower energy photons are absorbed by materials with lower band gaps (NIR). The bandgap of different semiconductors frequently used in thin-film multijunction photovoltaic devices is reported in Figure 1. [14]

Multijunction solar cells surely represent an efficient way to push photovoltaic devices beyond the Shockley-Queisser limit and set a new theoretical limit (around 86% for an infinite number of junctions) for conversion efficiency. However, those devices are not commercially viable due to the great economic resources needed for their fabrication into thin films. Very sophisticated techniques as molecular beam epitaxy (MBE) or metal-organic chemical vapor deposition (MOCVD) need to be employed to grow multilayer structures made of materials belonging to the III-IV-V groups (Ga, As, Ge, In and so on). [16] Such issue leads to the use of multijunction photovoltaic devices in fields where the cost is not a limiting factor, such as in aerospace, where they are used because of the high efficiency and their low weight. [1,14]

Moreover, considering the design of multijunction devices, one assumes the different absorbers as stacked one on top of the other. This of course limits all of them to work in given conditions set to maximize efficiency of the structure as a whole, being the materials optically and electrically connected in series. [14,17] However, according to Brown et al. [14] it has been shown that the incoming sunlight can be separated in bands with a narrower spectral distribution. This would lead to higher efficiencies as each band would be absorbed by each specific solar cell, thus avoiding thermalization losses. Despite the great number of advantages of such multilayer structure, the need for the cells to be grown on separate substrates would cause it to be too expensive. [14,18]

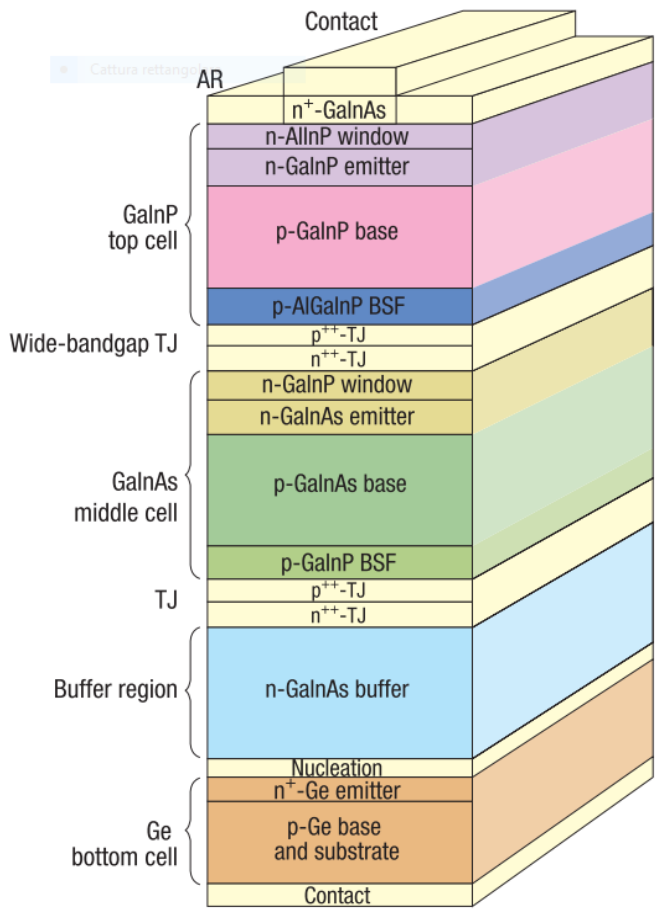
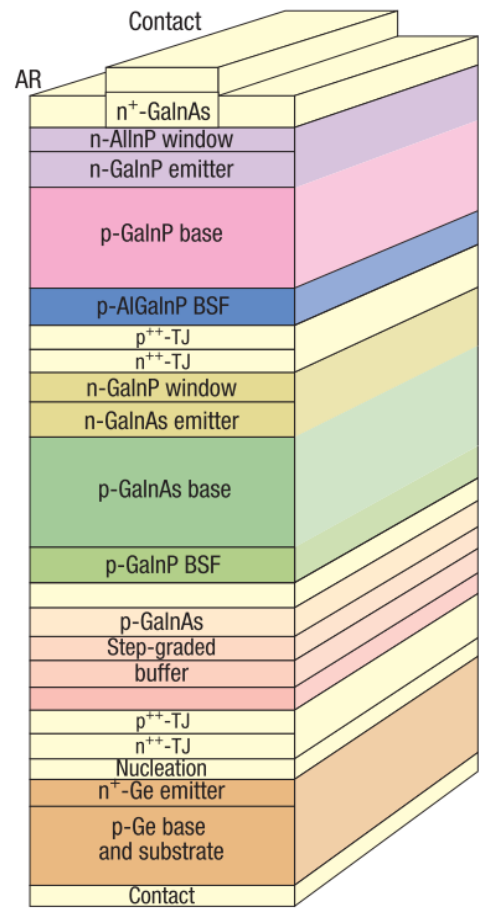

*Figure 2 Schematic cross-section representation of three-junction GaInP/GaInAs/Ge cell configurations: (a) lattice-matched and (b) metamorphic. (AR: antireflection coating; BSF: black-surface field; TJ: tunnel junction). Reprinted from [2]*

The workhorse of this category of solar cells, instead, is the lattice-matched three-junction GaInP/GaInAs/Ge cell (Figure 2). [2,19] The working range of this kind of devices extends from about 300 nm to around 1800 nm, stretching in the NIR beyond the limit of silicon based solar cells. [17,19] Such devices already show great efficiencies to date (>40%) as reported in NREL Best Research-Cell Efficiency chart. [20] In 2005, Yamaguchi et al. [19] reported technological improvements, which led conversion efficiency of InGaP/InGaAs/Ge multijunction devices to 31-32% (AM1.5) and 37.4 % in concentrated sunlight condition. In their paper they estimated that conversion efficiency would reach 40% sooner than expected, thanks to effortless research in the field [19,21]. Moreover, they stated that 4-junction devices obtained as a combination of a top cell with $E_g$

= 2.0 eV, a GaAs second-layer cell, a third-layer cell material with an $E_g$ of 1.05 eV, and a Ge bottom cell on a Ge lattice matched substrate has the potential to reach over 45% under 500 suns (AM 1.5) condition. [19]

Lattice matching is a necessary condition to obtain optical transparency and maximum current conductivity between the top and bottom cells in monolithic devices. This results in the choice of layers having similar crystal, or lattice, structure. In fact, mismatches in the crystal lattice constants between the layers originate defects and dislocations, which act as centers for recombination of carriers, ultimately degrading performance of the device by decreasing open circuit voltage, short-circuit current density and fill factor. [17]

Besides III-V devices, many other materials have been proposed as absorbers in lattice-matched multijunction solar cells. According to the plot proposed by Zhang et al. [22] there exists the possibility to fabricate junctions covering the whole solar spectrum form 3 eV to 0.4 eV by making use of II-VI and III-V quaternary alloys, still mantaining the lattice-matched condition. As suggested, such possibility opens up to reaching ultrahigh efficiencies, thanks to the fabrication of devices with a larger number of junctions.

Nonetheless, the solar cells achieving the greatest efficiencies are those exploiting metamorphic multijunction designs (Figure 2). [2] Metamorphic multijunction devices feature metamorphic buffer layers (graded semiconductor composition) in which dislocations are allowed to form. This leads to gradual relaxation of the crystal structure, resulting in a new, larger lattice constant, which will in turn be used as a new substrate for the growth of high-quality semiconductors with a larger crystal lattice constant. [2] As pointed out by King et al. [23] such devices provide an unprecedented degree of freedom in solar cell design, by allowing flexibility in band gap selection, unconstrained by the lattice constant of common substrates such as Ge, GaAs, Si, InP, etc. [23]

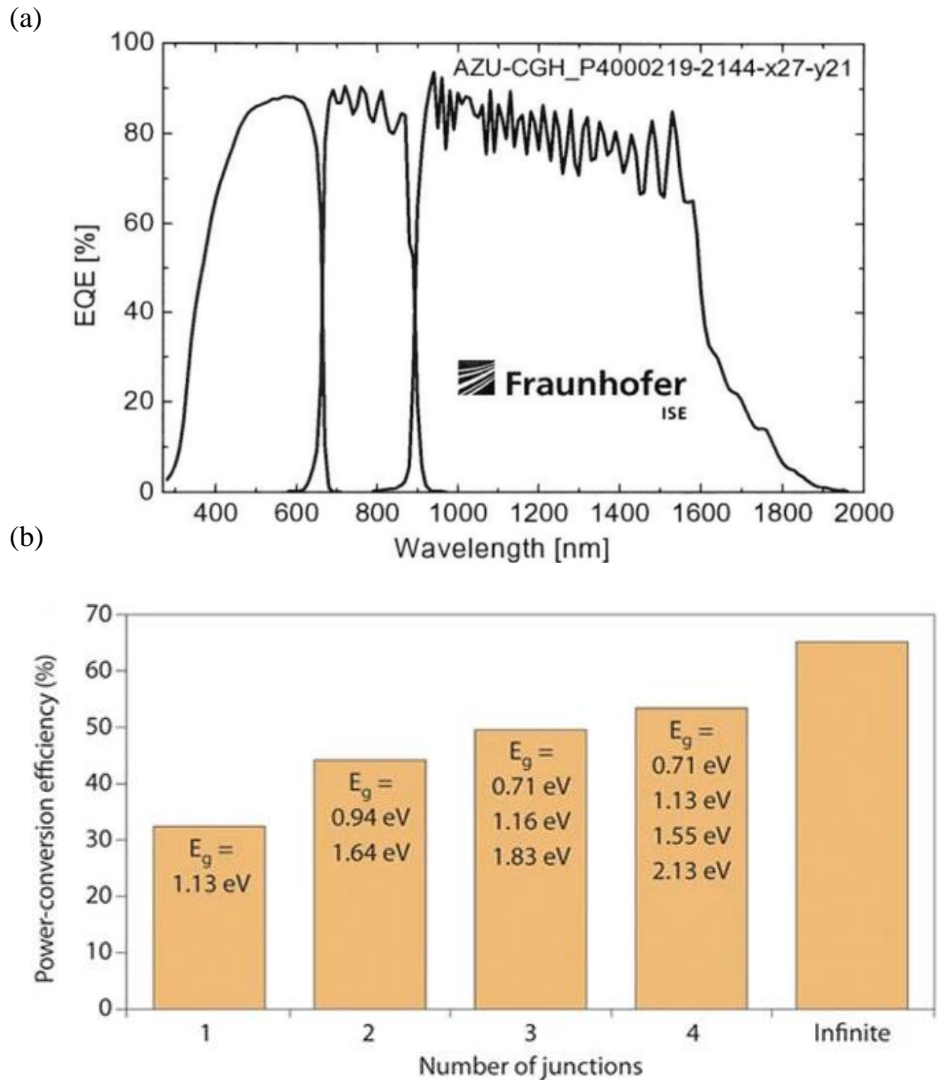

*Figure 3 (a) External quantum efficiency (EQE) of an In0.50Ga0.50P top cell, In0.01Ga0.99As middle cell and a Ge bottom cell fabricated by the company AZUR SPACE. Reprinted from [24] (b) Maximum power conversion efficiency that a hypothetical n-junction solar cell can achieve for AM 1.5 condition. Optimal bandgap as a function of the number of junctions is reported. Reprinted from [21]*

Metamorphic three-junction cells have been fabricated in the lab with high-indium-content Ga0.44In0.56P top subcells, and Ga0.92In0.08As middle subcells, at a lattice mismatch of 0.5% with respect to the germanium substrate, which also serves as the third and bottom subcell. As reported by works performed in several labs around the world, the germanium third subcell is replaced with a metamorphic GaInAs subcell that has a bandgap of around 1 eV, grown in an inverted configuration with a transparent graded buffer region. [2] This type of metamorphic cell design has given the best performance (33.8% efficiency) under unconcentrated sunlight in cells developed by the National Renewable Energy Laboratory. [2] However, efforts on achieving metamorphic devices with higher efficiencies (>40%) have been reported by companies leaders in the field (Figure 3). [24]

As previously mentioned, even if III-V multi-junction solar cells are those reporting the highest efficiencies, lowering their fabrication costs remains of utmost importance for their broad commercialization beyond the aerospace field, as it is for mainstream Si-based photovoltaics. However, as pointed out by Todorov et al. [18], developing innovative and cheaper multi-junction device structures can only be a long-term possibility, yet a much quicker and practical solution could be to upgrade large-volume technologies by integrating tandem devices on top. In this way, new technologies will take advantage of an already existing infrastructure, ultimately leading to a drop in their fabrication costs. [18] Nonetheless, the achievement of grid parity is still far, and it is not clear if it will ever be reached.

One of the main issues that scientists need to address when developing a tandem device is process compatibility. This is a major restriction during the fabrication process since bottom layers need to withstand the processing temperatures of the upcoming layers. One way to address the issue would be to bond two already complete devices, despite it being hardly feasible on commercial devices. Additional challenges arise from the necessity to introduce optimal recombination layers or tunnel junctions to reduce as much as possible parasitic absorption and series resistance. [18]

One of the main challenges regards the fabrication of junctions with bandgaps close to those predicted by theoretical assumptions (Figure 3) to obtain the maximum possible efficiencies out of the 3-, 4-, n-junction devices. The sub-cells need to obey to physical conditions mainly related to pairing with solar spectrum and current matching. This is extremely demanding in terms of materials science, which plays a key role in researching the most cost-effective procedure or materials to do so. The candidates to fulfill this purpose now include CdTe, CdSe, II-IV alloys, chalcogenide, and chalcopyrite materials. For a complete review on the status of different material sets under study for the fabrication of innovative multi-junction devices and prospects for commercial tandem solar cells refer to [18].

## 2.2 Organic-Inorganic Perovskite solar cells

Among the many materials showing photovoltaic behavior, those identified as the most promising candidates belong to the category of organic semiconductors and halide perovskites. Nowadays, solar devices fabricated by making use of such materials show conversion efficiencies comparable with those of well-established, long-lasting technologies based on Silicon. Moreover, fabrication is carried out exploiting wet chemistry processes, which have the advantage of being easy and straightforward and allow the deposition of the materials also on flexible substrates. However, besides showing some atmospheric stability issues which limit their large scale industrial production [25], those devices still focus on harvesting the visible part of the solar spectrum, with very few devices stretching to the nearest infrared, without being optimized for its absorption. An extensive review on the degradation routes (i.e. moisture, oxygen, UV light, thermal and electrical stresses) and stability investigations published in literature and pertaining large area perovskite modules fabricated both on rigid and flexible substrates can be found elsewhere. [25] According to Castro et al. [25] to improve lifetime of perovskite solar devices, especially regarding large modules, new materials and interfaces, and cells architectures which are inherently more stable must be developed, together with more effective encapsulation methodologies [25].

In just a few years, halide perovskites have achieved the most rapid efficiency growth among all photovoltaic materials (Figure 4). [25] The always growing attraction towards those materials also results from their countless possibilities for tandem applications, thanks to their tunable optoelectronic properties and low processing temperatures (lower than any other high-performance absorber). [26] Perovskite solar devices are characterized by high open-circuit voltage which they can generate under full sun illumination. As previously stated, sunlight is characterized by a broad spectrum, so that photons with lower energy than the material's bandgap are not absorbed, while those with greater energy relax to the bottom of the conduction band, losing excess energy in the form of heat. [26] The open-circuit voltage is the maximum voltage that the solar cell can generate, and it reflects the maximum energy that can be extracted from any absorbed photon. Consequently, it can be considered a simple measure of the fundamental energy loss in the solar device. [26]

Thanks to their processing conditions (e.g. low temperature) and countless possibilities for integration with other pre-made devices, solar cells based on perovskite absorbers promise to break the paradigm by combining

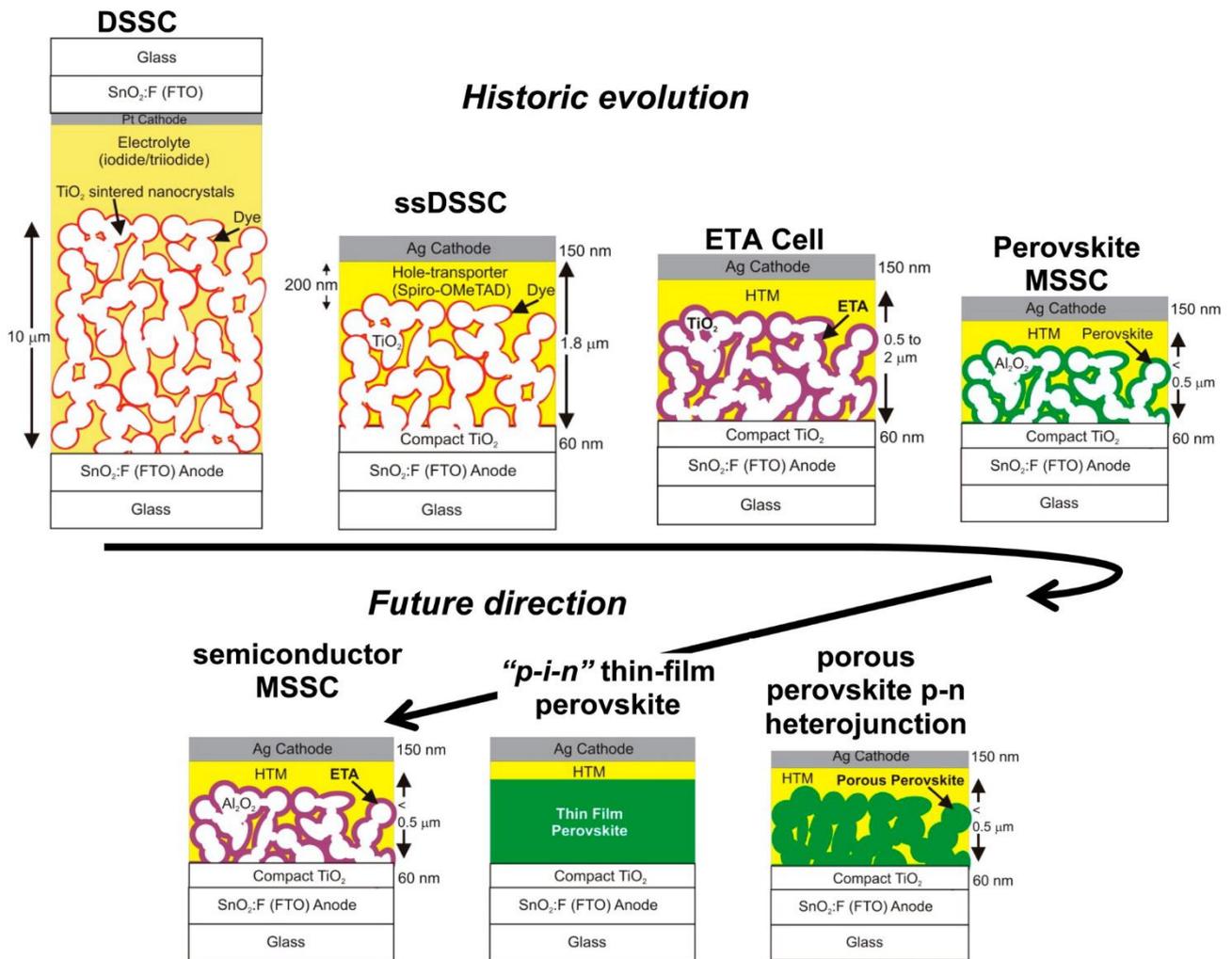

*Figure 4 Historic evolution of organic-inorganic perovskite based solar cells, starting from dye-sensitized solar cells (DSSC) to meso-superstructured perovskite solar cells (MSSC). Three possible future directions for the perovskite technology are illustrated: (i) porous perovskite distributed p−n heterojunction solar cells, (ii) thin-film p−i−n perovskite solar cells where no porosity is required and the device takes on the structure of an intrinsic thin perovskite film sandwiched between p- and n- type charge-extracting contacts, or (iii) semiconductor MSSCs, where any solution-processed semiconductor can be structured by the porous scaffold to deliver a MSSC. Reprinted from [26]*

both low cost and high efficiency. [26] Among the first high-efficiency devices, those fabricated with $CH_3NH_3PbI_3$ show a relatively high bandgap of 1.55 eV, but it can be easily tuned across a wide range through multiple cations substitutions, in order to increase or decrease the bandgap itself. [18] For example, bandgaps of 1 eV can be achieved by substituting Pb ions with Sn, while substitutions of I with Br will lead to 2.2 eV bandgap. [18]

Even if the optimization of low band gap devices is currently a hot topic ongoing in the perovskite research field, it already opened up a door to the fabrication of all-perovskite multi-junction devices. [18] According to Zhao et al. [27] metal-halide perovskite-only tandem solar cells represent an attractive choice for next generation photovoltaics, but their progress has been hindered by the lack of high performance low-bandgap perovskite devices. State of the art $CH_3NH_3PbX_3$ (where X = F, Cl, Br, or I) perovskites show poor light absorption over 780nm because of their bandgap of 1.55 eV. [28] However, considering that lead (Pb) and tin (Sn) halide perovskites are characterized by fine bandgap tunability, those can be employed as bottom cells in tandem devices. [27] Bandgap of mixed Sn and Pb iodide perovskites can be tuned from 1.17 eV (50% Sn and 50% Pb) to 1,58 eV (pure Pb). [28] Hao et al. [28] reported a mixed perovskite device ($CH_3NH_3Sn_{1-x}Pb_xI_3$) with absorption extending to 1050 nm (1.17 eV) as reported in Figure 5. Such device showed promising short-circuit photocurrent density of 20.64 mA/cm$^2$ when coupled with mesoporous TiO2 electrode and spiro-OMeTAD hole transport layer under simulated full sunlight of 100 mW/cm$^2$. The tunability of the bandgap results from the distortion of the crystal lattice of the material, due to the presence of Sn impurities, which influences charge-transport properties. [28] As a matter of fact, the $CH_3NH_3Sn_{0.5}Pb_{0.5}I_3$ originally reported by Ogomi et al. [29] resulted in an anomalous behavior for what regards bandgap tunability. [29] Mixed compositions showed a trend as a function of Sn impurities which was not linear, with intermediate compounds having the smallest bandgaps as shown in Figure 5. [28,29] Spectral response of the devices is additionally evidenced by the IPCE of devices fabricated with different ratios. [28]

(a)  (b)

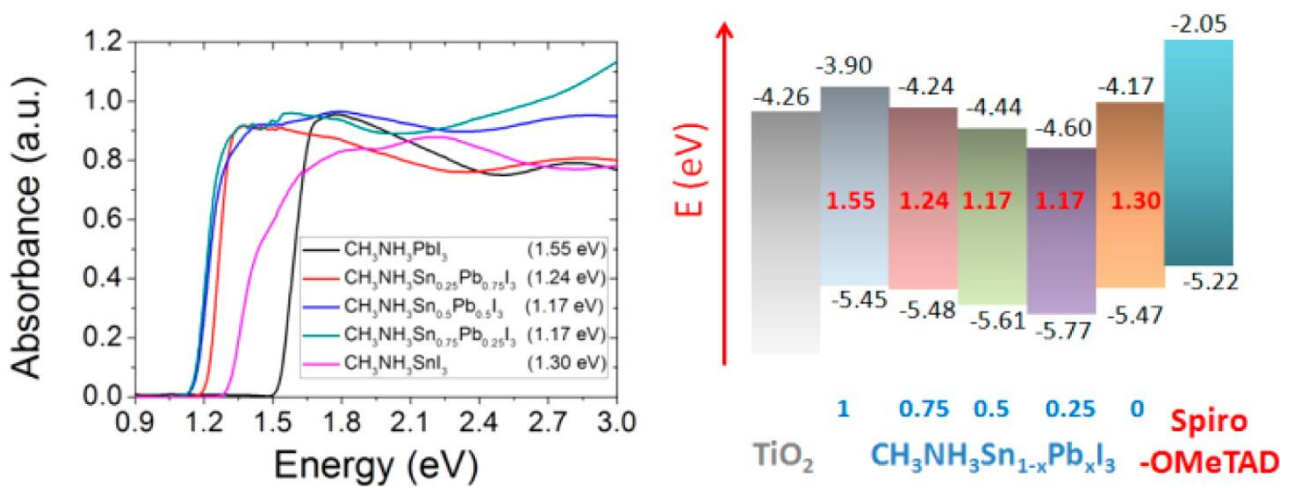

*Figure 5 (a) Absorption spectra of mixed Pb/Sn organic perovskite devices. (b) Energy level diagram of $CH_3NH_3Sn_{1–x}Pb_xI_3$ solid solution perovskites. Adapted from [28]*

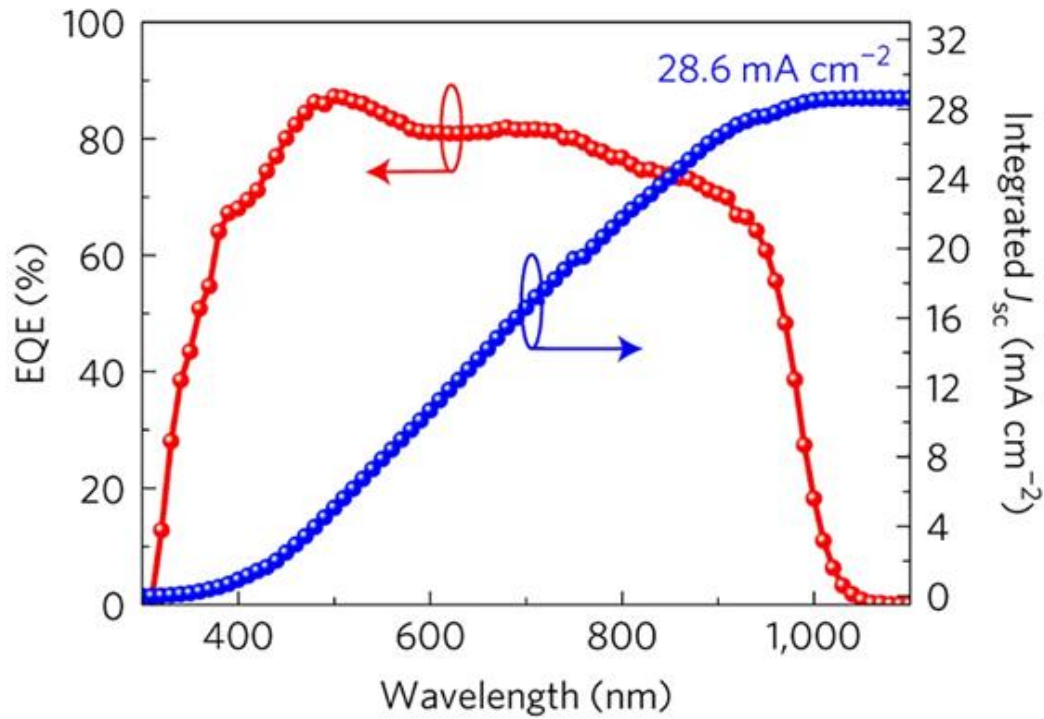

*Figure 6 Figure 7 EQE spectrum and integrated current density ($J_{SC}$). The integrated $J_{sc}$ over a 100 mW cm$^{-2}$ AM1.5G solar spectrum is 28.6 mA cm$^{-2}$. Reprinted from [27]*

All-perovskite multijunction devices have already been demonstrated with encouraging efficiencies. Eperon at al. [30] reported the fabrication of four- and two-terminal perovskite devices with ideally matched bandgaps. It must be recalled that a multijunction device showing the highest possible efficiency would require a bottom cell with a bandgap ranging from 0.9 eV to 1.2 eV and a top cell with a bandgap ranging from 1.7 eV to 1.9 eV. [27,30] In 2016, the solar cell based on $FA_{0.75}Cs_{0.25}Pb_{0.5}Sn_{0.5}I_3$ developed by Eperon et al. [30] and showing 14.8% efficiency with a $V_{OC}$ of 0.83 V was combined with a 1.8 eV $FA_{0.83}Cs_{0.17}Pb(I_{0.5}Br_{0.5})_3$ cell to demonstrate a monolithic all-perovskite two-terminal tandem solar device, which was current-matched and efficient (17% PCE on small areas) with $V_{OC} >$ 1.65 V. [30] A four terminal 20.3% efficient small-area all-perovskite tandem device was also reported by the same group. [30]

Those numbers were topped by multijunction devices fabricated by Zhao et al. [27], exploiting the anomalous bandgap tunability of mixed Pb and Sn based perovskites. In fact, in 2017, the group reported an all-perovskite 4-terminal tandem solar cell with a maximum PCE of 21.2%. [27] Such record was achieved thanks to the fabrication of a relatively thick (around 600nm) absorber layer (mixed Pb and Sn perovskite) with large grains and long carrier lifetimes. [27] Having a thicker layer can make it up for the low spectral response in the 700nm to 900nm spectral range and allow for greater EQE (Figure 6). [27]

Despite the encouraging results already achieved, many efforts from researchers still focus on the integration of wide-bandgap perovskite solar cells with Si, Copper Indium Gallium Selenide, and polymers to be used as bottom cells. [27] However, all-perovskite tandem devices would be advantageous in terms of manufacturing costs and easy processing. [27]

## 2.3 Organic Photovoltaics

Together with organic-inorganic perovskite materials, another alternative for the development of low-cost devices characterized by moderate efficiencies is represented by printable organic photovoltaic devices exploiting active layers based on semiconducting polymers and small organic molecules. [16,31,32] The use of these materials, which are inherently inexpensive, combined with their compatibility with low-temperature processing and the possibility of high through-put production processes, allows the fabrication of flexible, lightweight devices, whose main characteristic is the presence of strongly bound exciton states (Frenkel excitons). [16,31,33–35] Excitons influence most processes in the devices, including strong absorption and photocurrent generation. [16,34] An excited bound electron-hole state is generated by light absorption and subsequently travels in the material, thanks to the built-in electric field, towards a donor/acceptor interface where it will be split into free charges, which can be collected at electrodes (Figure 7). [16,34] Throughout the years, many studies focused on donor/acceptor interface architecture [32,33,36,37]. The road to efficiency maximization winds through optimization of nanostructured bulk heterojunctions between the absorber and the acceptor to decrease excitons travel distances in the active layer and favor charge separation and collection of free charges. Additionally, thickness reduction of the active layers allows the fabrication of semi-transparent devices paving the road to integration of photovoltaics into buildings, cars and so on. [16,35]

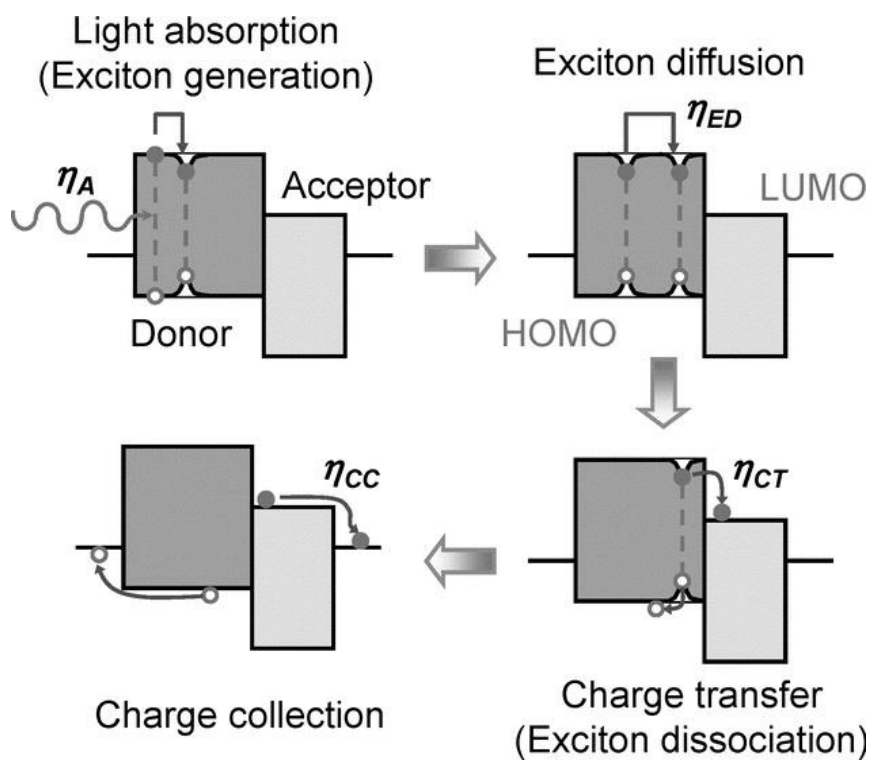

*Figure 8 Schematic diagram of the steps of photovoltaic process in an organic donor-acceptor heterojunction. Light absorption generates an exciton which diffuses towards donor-acceptor interface. Charge transfer occurs at the interface and causes the exciton to dissociate into free charges which will separately diffuse in the HOMO of the donor material (hole) and the LUMO of the acceptor material (electron) until reaching electrodes. Reprinted from [31]*

However, an additional boost to efficiency could derive from the implementation of devices containing layers of NIR-absorbing organic molecules / polymers acting as active layers or in tandem devices in a fashion similar to perovskites. [37] Nonetheless, synthesizing organic molecules with long diffusion length, high carrier mobilities and with an optical bandgap engineered to achieve absorption in the NIR has resulted challenging. [34] Moreover, tandem devices between organic photovoltaics and perovskites themselves constitute a valid possibility thanks to the compatibility of processing techniques i.e. spin-coating, bar coating, low processing temperatures. Combining these technologies together with new organic- or perovskite-based NIR absorbing materials could lead to great results in terms of efficiency.

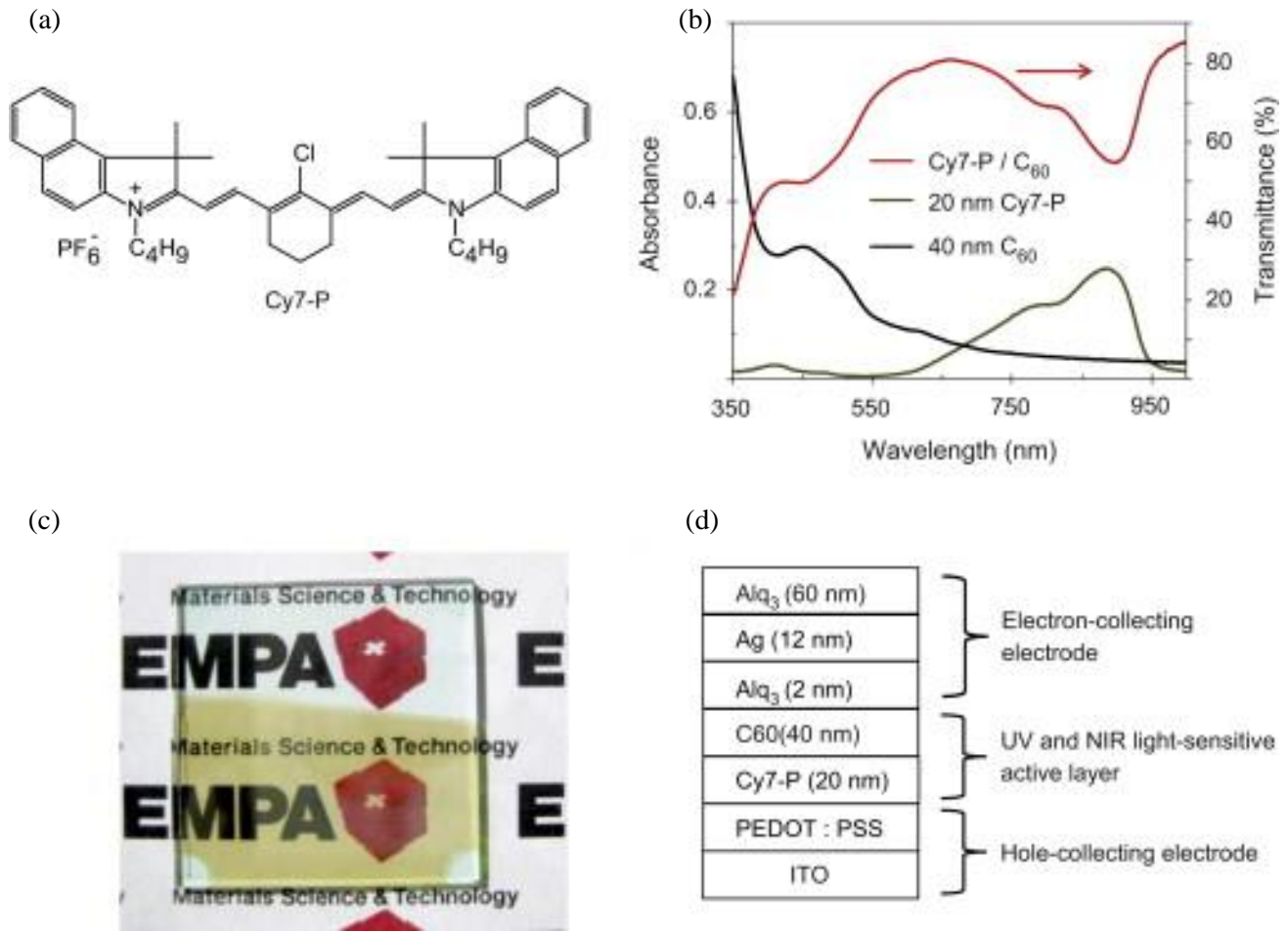

*Figure 9 (a) Molecular structure of Cy7-P. (b) Absorbance spectra of separate C60 and Cy7-P films, and transmittance spectrum of the corresponding bilayer film. (c) Picture of a Cy7-P and C60 coated glass subsrtrate. (d) Schematic of the device architecture for a semitransparent Cy7-P/C60 solar cell. Adapted from [35]*

In 2013 H. Zhang et al. [35] reported the use of a cheap, commercially available, solution-processed hepthamethine dye (Cy7) with terminal benzidol moieties as electron donor (Figure 8) coupled with C60 as electron acceptor. [35] By making use of such molecule, the group was able to fabricate a solar cell with 65% transmittance in the visible with spectral response both in the UV and NIR. [35] The best performance cell ($\eta$ = 1.5%) was achieved for a Cy7 thickness of 20 nm, showing $J_{sc}$=6.2 mA/cm2, VOC = 0.38 V, FF = 58.4%. [35] According to the authors, those results simply introduce cyanine dyes as a suitable class of materials for the fabrication of semitransparent organic solar cells, whose spectral response in the NIR could reach 1000nm and beyond, depending on the molecules. [35]

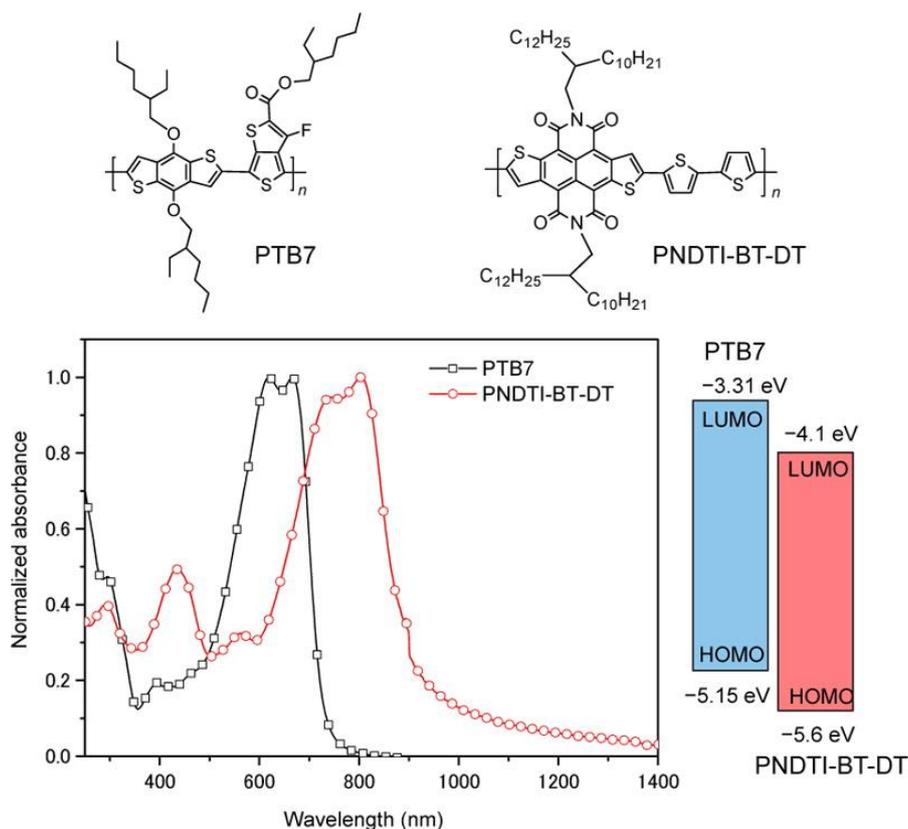

*Figure 10 Chemical structures of PTB7 and PNDTI-BT-DT, together with their absorption spectra in film and energy level diagrams. Reprinted from* [38]

According to Winder and coworkers [39] absorption of certain conjugated polymers can be shifted to the red part of the visible or even in the NIR, but this would lead to lower absorption in the blue-green region. To overcome this problem, as they suggest, organic dye molecules or wide band gap polymers can be used as sensitizers to be added to the donor/acceptor polymeric blend, in order to absorb in this high energy region and ultimately carpet the sun spectrum. [39] As an example it has been shown that PTPTB can be sensitized with MDMO-PPV and nile red. Devices made with such combination show photocurrents peaking at 550nm which is the absorption maximum of the dye molecule, but the photocurrent response is observed also in the 600 to 750nm region, which is the absorption range of PTPTB polymer. [39]

In 2014, Zhou et al. [38] reported a novel polymer based on NDTI and bithiophene with extended absorption in the NIR up to 900nm and high electron mobilities. [38] Such polymer was combined with PTB7 in an all-polymer solar cell, which showed high photoresponse in the NIR (Figure 8). [38]

Another way towards NIR-absorbing organic photovoltaics according to literature goes through the fabrication of devices made of copolymers. The synthesis of p-type Donor / Acceptor copolymers has been proven to be an interesting way to achieve separate control of LUMO and HOMO levels and successfully increase EQE and absorbance in the NIR. [40] In 2004, X. Wang et al. [41] fabricated a device by making use of a novel copolymer (APFO-Green1) able to absorb light in the 300 < λ <500 and 650 < λ <100 nm wavelength range. [41] The polymer was then coupled with different acceptors and showed photovoltaic behavior in the infrared up to 1000nm. [41] Such result highlights the possibility of carpeting the solar spectrum with visible and NIR absorbing copolymer building blocks.

An extensive list of novel materials to achieve infrared absorption and high visible transparency has been published in a recent review by Chang et al. [40]. Beside mentioning polymeric materials with low bandgap that have recently been investigated for such purpose, the authors also shine light on the different techniques and designs exploitable to have visibly transparent and highly efficient photovoltaic devices with fully organic active materials.

## 2.6 Infrared quantum dot solar cells

Colloidal quantum dot solar cells introduce interesting possibilities as they combine the properties of semiconductor materials and the low-cost processability of organic materials. [42–44] The energy gap of semiconductor nanocrystals is characterized by the quantum size effect. Due to the confinement of the electrons in nano-scale objects, quantum effects become more significant causing the energy of the system to be quantized. Limiting the discussion to the case of spherical nanocrystals, the energy gap can be written as [45,46]:

$$E_g = E_{g,bulk} + \frac{\hbar^2 \pi^2}{2m_e a^2} + \frac{\hbar^2 \pi^2}{2m_h a^2} - 1.8 \frac{\hbar^2}{\mu r_B a} \qquad (2.1)$$

The first term $E_{g,bulk}$ is the energy gap of the corresponding bulk material. The second term is the quantization energy related to the zero-point energy of electrons, with $\hbar = 6.58 \times 10^{-16} eVs$ being the reduced Planck constant, $m_e$ the electron effective mass, and $a$ the quantum dot diameter. The third term is the quantization energy related to the zero-point energy of holes, with $m_h$ the hole effective mass. The fourth term is the contribution due to the Coulomb attraction between electrons and holes, with $\mu$ the reduced mass of the exciton, defined as $1/\mu = (1/m_e + 1/m_h)$, $r_B$ the Bohr radius of the exciton, defined as $r_B = \hbar^2 \varepsilon^2 / \mu e^2$ (where $\varepsilon$ is the dielectric constant of the semiconductor and $e$ the electron charge).

As a straightforward consequence, the energy gap of a quantum dot is blue shifted with respect to the corresponding bulk material. For instance, the energy gap of bulk PbS is about 0.4 eV, while PbS quantum dots with a size of 3.6 nm show an energy gap at about 1.15 eV. [47] On the other hand, the advantage of quantum dots with respect to the corresponding bulk is the very high extinction coefficient near the band edge. [48,49] Thus, a reduced amount of material can be used to harvest Sun's irradiation, leading to thin photovoltaic devices.

Another issue for quantum dot based solar cells is the charge transport through the layers of material. While the bulk material shows band-like transport, it has been reported that electron transport in PbSe quantum dot films show variable-range-hopping or nearest-neighbour-hopping behaviour depending upon temperature. [50]

To reach the infrared region with colloidal quantum solar cells, the workhorses are materials like PbS and PbSe nanocrystals, due to the low energy gap of the corresponding bulk. [51–54] Unfortunately, such materials are toxic and more environmentally friendly materials have been investigated. For example, solar cells based on AgBiSe$_2$ [55], on Ag$_2$S [56], Cu$_2$GeS$_3$/InP core/shell [57] colloidal nanocrystals have been reported.

## 2.7 Upconverting solar cells

Upconversion based solar cells represent a different type of infrared solar cell. In these devices, the active material does not have the task of absorbing the infrared part of the Sun's irradiation, which is in fact absorbed by an additional layer. Such energy is then transferred in the range in which the active material is able to absorb light via a multi-photon relaxation process.

A common strategy employs Ytterbium ions and Erbium ions and involves the excitation of Ytterbium at 980 nm and, after energy transfer processes between Yb$^{3+}$ and Er$^{3+}$, the emission in the green and the red regions. [58]

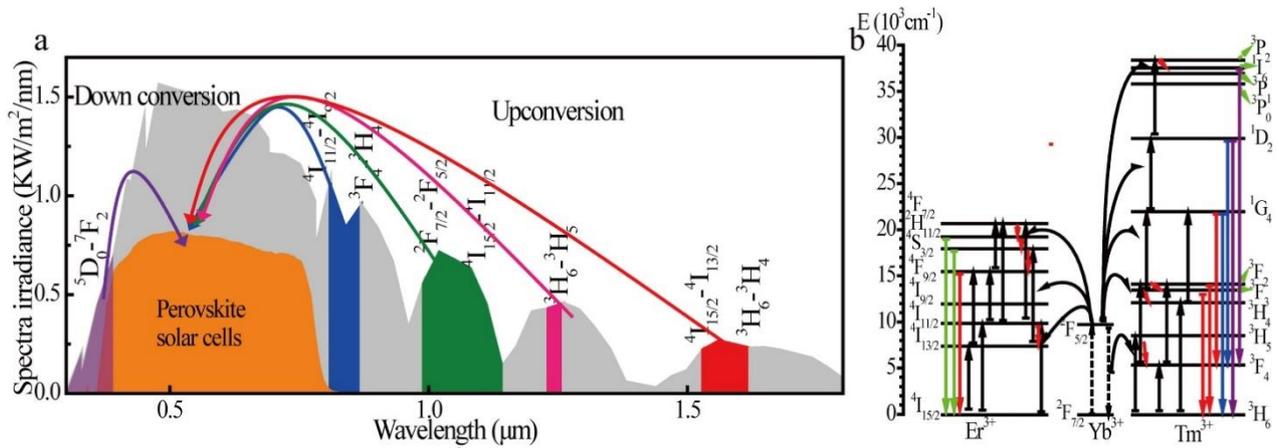

*Figure 11 (a) AM 1.5G spectrum and fraction absorbed by organic-inorganic perovskite solar cells. The regions of the spectrum that can be accessed through upconverting and downconverting processes are also shown. (b) Schematics of the energy levels involved in the upconverting and downconvertin processes occurring in the materials ($Yb^{3+}$, $Er^{3+}$ and $Tm^{3+}$). The path that is interesting for upconversion is the excitation of $Yb^{3+}$ at 980 nm, energy transfer to the Erbium ion, and emission by Erbium ion in the green and in the red. Adapted by [59]*

NaYF$_4$ doped with $Yb^{3+}$ and $Er^{3+}$ is a very efficient up converter material absorbing photons in the infrared with an efficiency of about 50% and emitting light in the visible range. [60,61]

Another interesting system involves a layer of quantum dots below the $Er^{3+}$-based upconverting layer, in which the quantum dots absorb in a broad range in the infrared and emit photons at about 1520 nm, resonant to the absorption of $Er^{3+}$. [62]

There are many works reporting the use of upconverting nanostructures for the enhancement of organic-inorganic perovskite solar cells. [63–67] Li et al. [59] were able to achieve full spectral response (UV-VIS-NIR) by introducing two different photoluminescent converting layers able to upconvert light to NIR and downconvert light to UV. [59] In a work from Zeng et al. [68], lanthanide-doped upconverting nanoparticles are used to sensitize all-inorganic Caesium lead halide perovskite CsPbX$_3$ (X = Br, Cl, and I) quantum dots, which have emerged as promising materials for applications including solar cells and LEDs and can only be excited by short-wavelength light with high power under 600nm. [68]

Near infrared-to-visible upconverting films can be also based on organic materials. For example, a system based on triplet-triplet annihilation including rubrene nanoparticles and Os(atpy)(tbbpy)Cl$^+$ has been coupled to organic-inorganic perovskite solar cells. [64]

Upconverting nanoparticles have been also employed to enhance the absorption of dye-sensitized solar cells by fabricating TiO$_2$/upconverting nanoparticles nanocomposite layers or by simply mixing the nanostructures with the dye molecules. [69–73] In 2018, Zhao et al. [74] reported upconverting CeO$_2$:Yb,Er@SiO$_2$@Ag double-shell electro-spun fibres showing high upconversion luminescence due to the amorphous silica coating and the surface plasmon resonance effect of Ag. [74] The nanocomposite was used as an assistant layer in DSSCs, resulting in increased photoelectric conversion efficiency and increased response in the infrared, up to 1000nm. [74]

An interesting work by De Wild et al. [75] reported a slight increase in power conversion efficiency by coupling a layer of upconverting nanoparticles to a thin film hydrogenated amorphous silicon (a-Si:H) solar cell. Such result was interpreted as a proof of principle for the use of upconversion materials coupled with thin film solar cells.

## 3. Plasmon-induced hot electron extraction

In the attempt of extending energy harvesting in the IR spectrum, the development of highly doped semiconductors nanostructures exhibiting plasmonic effects has recently brought new advantages. [76–81] So far, the most studied and commonly used plasmonic nanostructures have surely been Au and Ag nanospheres due to simple production techniques [81–85]. Noble metals such as Ag, Au or Cu have the peculiar ability of absorbing and scattering light at visible and UV wavelengths [64,86,87], combined with a good chemical and physical stability [1]. These properties have been exploited by people for centuries when metal nanospheres were used for fabricating colored glass [88]. Still, the use of noble metals for photovoltaic applications has many unsolved drawbacks. Precious metals such as Au and Ag are expensive even in bulk shapes. Therefore, the use of such materials on the large scale would not be economically viable. A physical limitation of these nano-objects stands in the complexity of tuning the frequency at which the phenomena of efficient absorption and scattering by the metal nanoparticles can occur [88]. This, in turn, resulted in a limited usage of plasmonic materials in self standing photovoltaic applications. In these fields, highly doped semiconductors experiencing plasmonic properties paved the way for the development of cheaper and promising alternatives.

## 3.1 Optical properties of plasmonic nanostructures

When light impinges on a metal-like medium, such as noble metals or highly doped semiconductors, the energy of the radiation can excite a resonant plasmonic oscillation i.e., a collective and coherent motion of charge carriers within the material. While on planar metal interfaces it gives rise to Surface Plasmon Polaritons (SPPs), that can easily be treated theoretically, in the case of metal nanostructures, a full theoretical description can be very cumbersome, even for simple, symmetric shapes. Being the plasmonic oscillations confined within the metal nanostructures, these are referred to as Localized Surface Plasmons (LSPs). The resonant frequency of these oscillations can be estimated with the Drude-Sommerfeld model, thereby assuming the free motion of electrons near the Fermi level. The dielectric function is then modelled as the linear combination of two terms [83]:

$$\varepsilon(\omega) = \varepsilon_{IB}(\omega) + \varepsilon_D(\omega) \qquad (3.1)$$

The first term includes the role of interband transitions from the d-band electrons, while the second is a Drude term which accounts for the near-free motion of conduction electrons at the Fermi level. In the limit of the dipole approximation the nanoparticles are much smaller than the radiation's wavelength and the field can be assumed to be uniform and slowly varying over the whole nanostructure. In this regime, the real and imaginary components of the Drude term can be defined from:

$$\varepsilon_D = 1 - \frac{\omega_p^2}{\omega^2 + i\gamma\omega} \qquad (3.2)$$

where $\omega_p$ is the Plasma Frequency, defined as follows:

$$\omega_p = \sqrt{\frac{ne^2}{m_{eff}}} \qquad (3.3)$$

being $m_{eff}$ the effective mass of the electron, $e$ the electron charge, and $n$ the charge density. [80] The Plasma Frequency is proportional to the energy needed to stimulate a collective oscillation of charges moving through the lattice of the plasmonic material, while $\gamma$ is a damping coefficient. Physically, it represents the duration of the gradually lost coherence between the individual oscillators. It therefore depends on electron-electron, electron-photon and electron-defects scattering events. [89] Depending on the size of the nanoparticles and on the mean free path travelled by the charges, scattering with impurities and trap states on the surface may become more significant. [85,89–92] This causes a weak but real dependence of the plasmonic resonance frequency of a nanosphere on its radius. [90] For almost free electrons in metals the dephasing rate is usually much lower than the radiation's frequency and its role in the definition of the Drude term of the dielectric permittivity can be neglected. [92] The equation 3.3 of the Plasma Frequency shows its dependence on the charge carriers' density. As it can vary greatly among different materials, it constitutes the main variable in determining the optical plasmonic properties of metal nanostructures. Metals such as Au, Ag or Cu have their resonance frequency in the visible or UV spectrum range (around 550-600nm for Au and Cu [93,94] and around 450 for Ag [95]) due to very high electron densities. By using highly doped semiconductors whose charge carriers' density is orders of magnitude lower [80], the resonance frequency can be shifted to the IR.

A complete description of the metal-light interaction is provided by Mie Scattering Theory (MST), although in the IR spectrum the problem can be greatly simplified. In the limit of the dipole approximation, the Quasi-Static Theory (QST), which assumes the field to be locally uniform over the nanostructures, yields reliable results. This accurate yet strong assumption allows for a simpler application of MST for a metal nanosphere. Under the impingement of a monochromatic plane wave a fraction of the electric field will be scattered back into the surrounding medium where it will interfere with the incoming one. By assuming the plane wave to be linearly polarized along the z-axis and adopting polar coordinates the electric fields inside ($E_1$) and outside ($E_2$) the nanosphere can be written as:

$$E_1 = E_0 \frac{3\varepsilon_2}{\varepsilon_1 + 2\varepsilon_2} (cos\theta\, n_r - sin\theta\, n_\theta) = E_0 \frac{3\varepsilon_2}{\varepsilon_1 + 2\varepsilon_2} u_z$$

$$E_2 = E_0 (cos\theta\, n_r - sin\theta\, n_\theta) + \frac{\varepsilon_1-\varepsilon_2}{\varepsilon_1+2\varepsilon_2} \frac{a^3}{r^3} E_0 (2cos\theta\, n_r + sin\theta\, n_\theta) \quad (3.4)$$

being $a$ the radius of the sphere. From these equations, many of the properties of plasmonic nano-objects can be understood. First, the electric field inside the sphere, $E_1$, is uniform across the bulk of the nanostructure. By minimizing or maximizing $E_1$, one can enhance or reduce the amount of scattered and absorbed radiation. By properly grouping the terms in the second addend of $E_2$, the polarizability $\alpha$ of the metal nanosphere can be defined as:

$$\alpha(\omega) = 4\pi\varepsilon_0 a^3 \frac{\varepsilon_1(\omega) - \varepsilon_2}{\varepsilon_1(\omega) + 2\varepsilon_2} \tag{3.5}$$

This quantity has a maximum obtained by imposing the Fröhlich condition [88,90]:

$$Re[\varepsilon_1(\omega_0)] = -2\varepsilon_2 \tag{3.6}$$

For it to hold, the assumption of a weak dependence of $\varepsilon_2$ on the frequency must be made. [90] By applying this condition to the Drude-Sommerfeld model for a metal nanosphere, the resonance frequency results to be [88]:

$$\omega_0 = \frac{\omega_p}{\sqrt{3}}$$

In such condition, the electric field is maximized both inside and outside the metal nanosphere thus promoting a local enhancement of the impinging field or Near Field Enhancement (NFE). In this model the resonance frequency $\omega_0$ can be tuned by modifying the dielectric constant of the environment $\varepsilon_2$ or by tuning the plasma frequency $\omega_p$, whose only independent variable is the charge carriers' density. In reality, the damping coefficient $\gamma$ is not null and it participates to the definition of the resonance frequency, especially for nanoparticles where the mean free path of the charge carriers becomes comparable with the particle dimensions. [90] In such cases, interactions with the surface become more significant and the damping coefficient $\gamma$ becomes a function of the radius $r$ of the particle.

$$\gamma(r) = \gamma_0 + \frac{A v_F}{r} \tag{3.7}$$

with $\gamma_0$ being the bulk damping coefficient and $v_F$ the Fermi velocity of the metal. [90] While different models result in different values for the variable A, the general dependence on 1/r is a shared result. The role of the radius of the sphere indicates the importance of the ratio between the area and the volume of the object. In the following paragraphs, some examples of different geometries with different aspect ratios will then be presented. The application of doped semiconductors instead of noble metals introduces a new degree of freedom for the tuning of $\omega_0$. In doped semiconductors, the density of carriers can be modified by adjusting dopant concentrations thereby making them good candidates for tunable absorbers. [90] The phenomenon of Localized Surface Plasmonic Resonances (LSPR) can then be exploited for the improvement of light harvesting efficiency of photovoltaic and sensing devices. By using materials with a density of carriers much lower than typical noble-metals, lower plasmonic resonance frequencies can be obtained. The dependence of the plasmonic resonance frequency of the nanoparticles on carriers' density, [85] as well as shape and size, allows for the engineering of a broad band absorption in the visible as well as in the infrared spectrum. Plasmonic nanostructures are then embedded inside the active layer of photovoltaic devices to increase optical absorption via three main methods: light trapping, energy transfer and plasmon induced charge separation (PICS) and transfer.

## 3.2 Light Trapping

In the attempt to reduce the cost of photovoltaic modules, producers and researchers have attempted to reduce the thickness of the absorbing layer. [96] Thin modules allow for the fabrication of flexible solar cells but lower the efficiency of the individual cells. Due to the reduced thickness, the optical path in the active layer and therefore the portion of light being absorbed are significantly reduced. [97] To increase the efficiency of thin solar cells it is mandatory to both reduce the amount of radiation reflected on the air-semiconductor interface and to maximize absorption in the active medium. [72,82,84,85,89,97–99] Most often, producers can reduce reflectivity by preparing textured surfaces which, in turn, present defects that act as recombination centres. [85] The application of plasmonic materials to enhance light trapping by forward and backward scattering of the incoming light is well established as a way of increasing efficiency of solar cells. [83,85,89,100–103] Light trapping is probably the most trivial use of plasmonics to enhance photo-conversion in semiconductors. The basic principle is the use of metal nanostructures to prevent light from escaping the active region so to increase the chance of every photon being absorbed. When metal nanospheres are deposited on the surface of a material they generate strong variations of the refractive index across the film causing

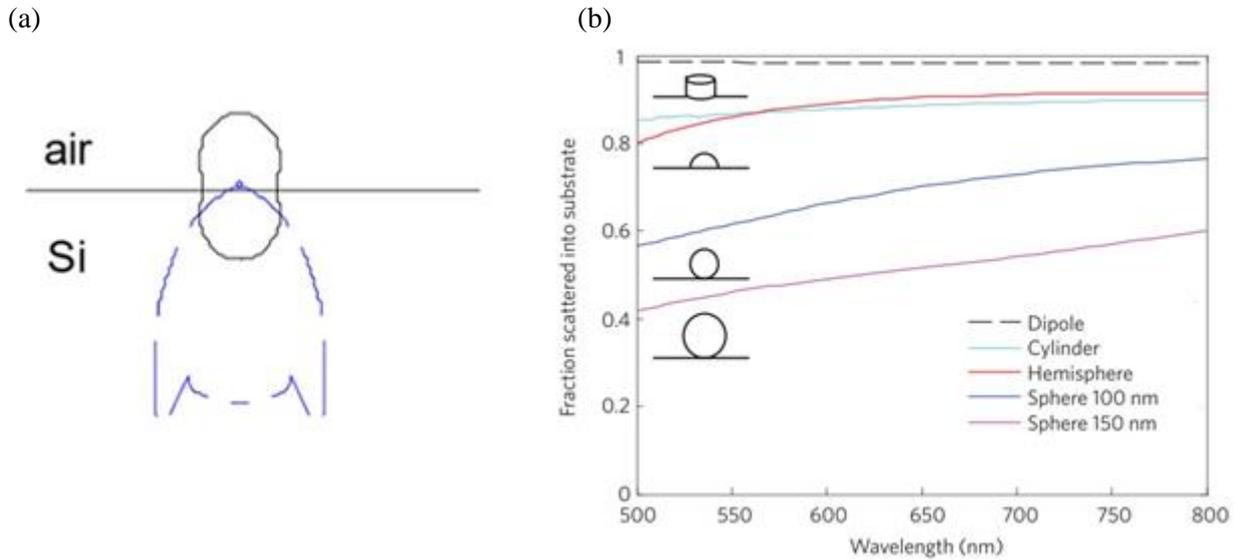

*Figure 12 (a) Radial distribution of the scattered field for a metal nano-particle deposited 20nm above the surface of the Silicon bulk in blue. In black, for comparison, the case for the nano-object in free space. Adapted from [85] (b) The share of scattered radiation, at different wavelengths, confined in the substrate depending on the shape and size of the nanoparticles. Adapted from [82]*

intense light scattering. [104] Kosei Ueno et al. [85] showed how most of the radiation scattered by a gold nanoparticle deposited just above the air-active layer interface is confined inside the material with the higher refractive index. The great asymmetry introduced in the scattered field is shown in Figure 11 (a). This generates a strong NFE inside the active layer upon illumination. [85] As metal nanospheres exhibit a cross section to the radiation much larger than their geometrical section [1], only a reduced number of nanoparticles is needed to effectively scatter a significant amount of the incoming light. [83,85] Harry a. Atwater et al. [82] analysed the dependence of the fraction of incident light scattered inside the silicon substrate on the shape and size of the metal nanostructure deposited on its surface. They showed that the best results were obtained for point-like dipoles due to their optimal near-field coupling with the substrate. [82] The plasmonic material is often deposited also on the metal-semiconductor interface preventing light from escaping the active region via backward scattering. [85] Au and Ag nanoparticles, deposited in single and double layers on the back side of the cell, form efficient back reflectors. They have been proven to significantly reduce total reflection and transmittance while greatly increasing diffused radiation. [85] The higher angle of propagation of the scattered light increases total internal reflection and the optical path travelled in the material. This enhances light trapping inside the medium allowing for an enhanced absorption of incident light. [72,100]

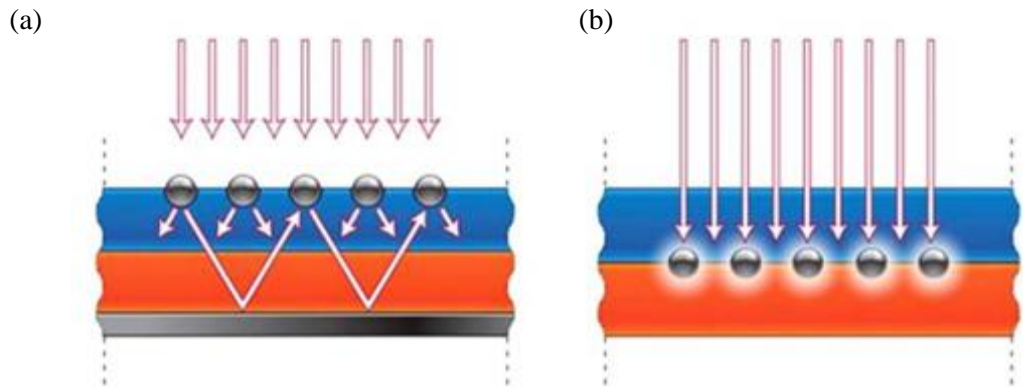

*Figure 13 Metal nanoparticles (a) deposited on the surface of the cell to improve light trapping and (b) embedded in the active layer for near field enhancement and absorption. Adapted from* [82]

Better light trapping allows for the engineering of thinner and cheaper cells, core principle for the development of flexible photovoltaics. [25] Using thinner cells also improves carriers' collection as the generated electron-hole pairs can be extracted more efficiently before recombination. [89,102,103] Since metal nanospheres deposited on the surface of the module can reflect part of the incoming radiation back into the air, they are often embedded inside the active layer of ultra-thin absorbers. [84,97,99] Figure 12 shows two possible configurations for arrays of metal nanospheres, either deposited on the surface of the cell or embedded in the active layer. By adjusting the spatial distribution of the embedded nanospheres in equally spaced arrays or by grouping them in specific arrangements it is possible to adjust the absorption bands of the material. [97] The response of the array is strongly dependent on the plasmonic coupling between the nanoparticles, which can be tuned by varying their reciprocal distance. A further improvement is obtained by inserting an oxide core in the metal nanoparticles. [99] The oxide core reduces the penetration of the field inside the nanosphere effectively limiting the absorption of the metal and enhancing the amount of scattered radiation. The different material also changes the effective refractive index of the nanosphere therefore shifting the plasmonic resonance frequency. [88] Despite the advantages of embedded nano-metals, stability issues remain. As the metal interacts with its surroundings, the defects on the interfaces may act as recombination centres for the generated charges. Stability can be ensured by creating an oxide shell on the nanosphere (Figure 13) at the cost of reducing the efficiency of NFE. As the scattered field drops quickly outside the metal surface, only ultra-thin oxide shells proved to be an efficient solution. [85,99]

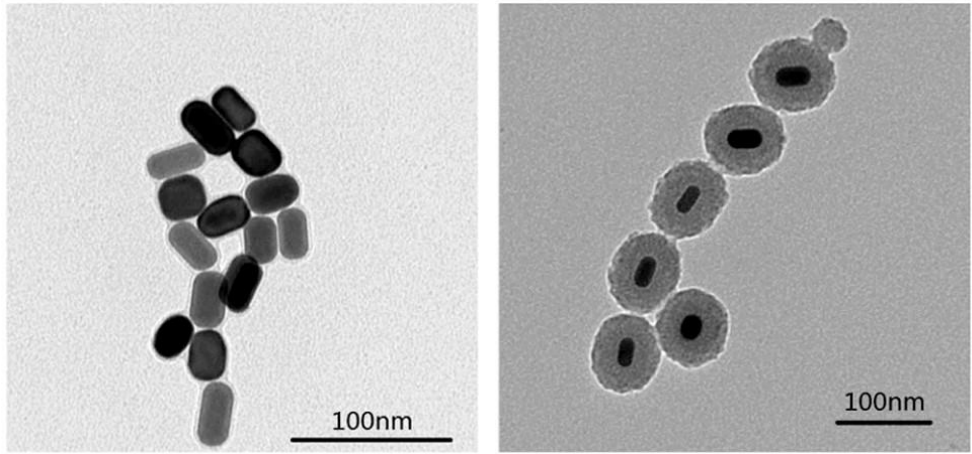

*Figure 14 Nano-rods with different aspect ratios covered in oxide shells for improving physical stability and tuning the plasmonic resonance frequency. Adapted from* [85]

## 3.3 Near field enhancement and plasmon induced resonant energy transfer

When light impinges on metal nanoparticles embedded in a semiconductor, its energy can be stored in the shape of LSPRs. These collective oscillations of charges can generate strong electric fields, up to orders of magnitude higher than the incoming one. [83] The energy of the field can then be transferred directly into the semiconductor through Plasmon Induced Resonant Energy Transfer (PIRET) where it generates an electron-hole pair. For the process to be efficient, the energy of the plasmon needs to overlap with the absorption bandgap of the semiconductor. Noble metals are often used due to their chemical and physical stability as well as the strong exhibited near-field enhancement. Previously, in equation 3.4 it has been shown how, under illumination from a specific resonance frequency $\omega_0$, the electric field outside a metal nanosphere can be strongly enhanced in its proximity, with a spatial decay proportional to $r^{-3}$. Due to the limited spatial extension of the generated enhanced electric field, the plasmonic nano-object needs to be deposited in the near proximity of the active layer. Since no direct transfer of charges occurs from the plasmonic material to the semiconductor, the PIRET is allowed even if an insulator is introduced between the metal and the active region (Figure 14). Metal nanostructures can therefore be covered by oxide shells that allow for a slight tuning of the plasmonic resonance frequency as well as improving stability issues. The oxide shells can in contrast increase the number of trap states increasing collision events of the oscillating charges. These cause random phase jumps in the

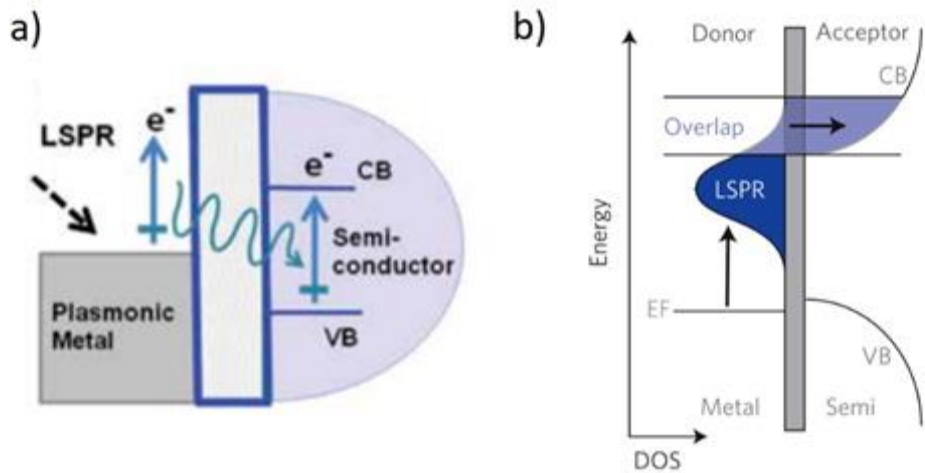

*Figure 15 (a) PIRET promoting the generation of electrons in the conduction band of a semiconductor nearby, across an insulating barrier placed between the plasmonic metal and the semiconductor. Adapted from [105] (b) The overlap between the tails of the distribution of hot electrons, generated through LSPR, and the conduction band of the semiconductor allow for the extraction of charges even with photons with insufficient energy to overcome the semiconductor's bandgap. Adapted from [92]*

harmonic motion of the electrons quickly damping the polarization of the nanoobject. As dephasing increases, coherence between the individual oscillating charges in the metal decreases and the possibility of energy transfer is progressively lost. [89]

For small Au nanospheres ($r < 15nm$), surface interactions become very significant, effectively increasing damping of the plasmon's coherence. For smaller structures, the charge population quickly degenerates into a hot carrier distribution and most of the energy is lost in the near-field or is absorbed. To increase RET efficiency, larger structures, such as spheres or rods, need to be employed so that dephasing times approach those typical of metal bulks. The phenomenon of surface plasmons can manifest also in the form of Surface Plasmon Polaritons (SPPs). These are surface electromagnetic waves, bound to the flat interface between a homogeneous metal and a homogeneous dielectric. SPPs are caused by the coupling between the electromagnetic radiation and the oscillation of free charges on the metal's surface. [88] Due to boundary conditions on the metal-dielectric interface, only SPPs with very specific wave vectors can be generated. The excitation of SPPs is then strongly dependent on the angle of incidence of the light and on its polarization. [88] Depending on the refractive indexes of the materials at the interface, only very specific modes of SPPs can be induced thereby limiting the width of the spectrum on which energy transfer through near field enhancement can occur. Also, being SPPs propagative modes, they cannot be generated on isolated nanoparticles and instead

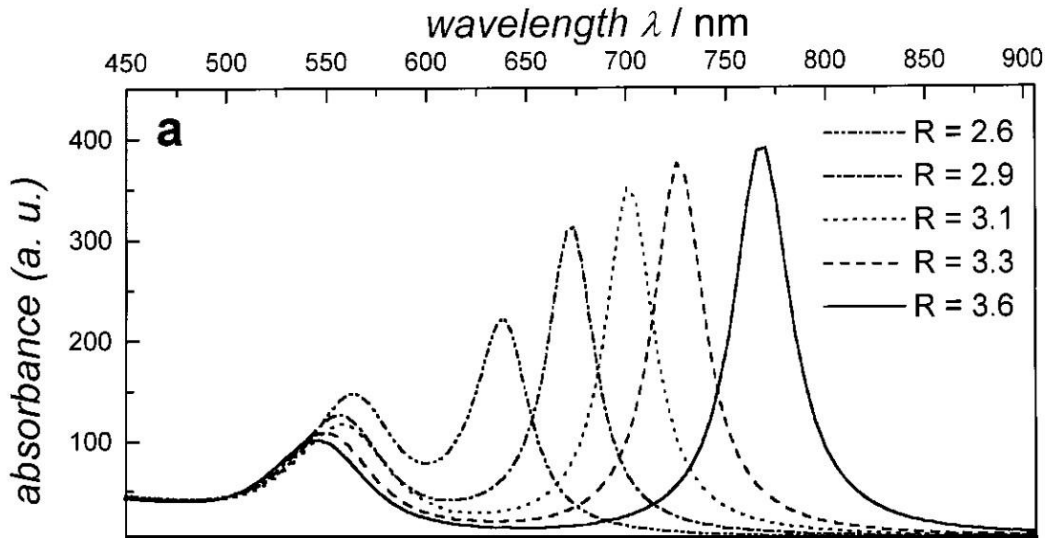

*Figure 16 Absorption spectra of metal nano-rods with varying aspect ratio R. While the high frequency mode, attributed to transversal modes, is subjected to only a small blue-shift, a second more prominent peak, attributed to longitudinal modes, progressively moves toward longer wavelengths. Adapted from* [90]

require precise geometries to be designed. Due to the many drawbacks of SPPs, most of the research about plasmonic-driven efficiency enhancements today focuses on LSPR. [85] The advantage in exploiting LSPR stands in the strong dependence of the plasmonic resonance frequency on the material and on the shape of the nano structures. In this scenario, highly doped semiconductors have shown a very promising role due to a much lower plasmonic resonance frequency in the NIR. Following the excitation, the energy of the plasmon is quickly shared among the oscillating charges in a distribution centred around the LSPR's resonance frequency. The high energy tails of such distribution, as shown in Figure 14, can overcome the energy barrier of the semiconductor even for plasmons generated at lower energy than the semiconductor's bandgap. PIRET can then effectively allow absorption at longer wavelengths widening the absorption band of the semiconductor. [83,92,102]

## 3.4 Aspect Ratio

Nanospheres are still the most common and simple structures used in plasmonic devices. [1,81,85] Despite them being the easiest shape to fabricate, the spherical symmetry limits the number of degrees of freedom that can be used to modify the position of the peak of the absorption spectra. More complex and less symmetric geometries instead allow for the extension of the absorption band into lower frequencies. [1,90,106] The loss

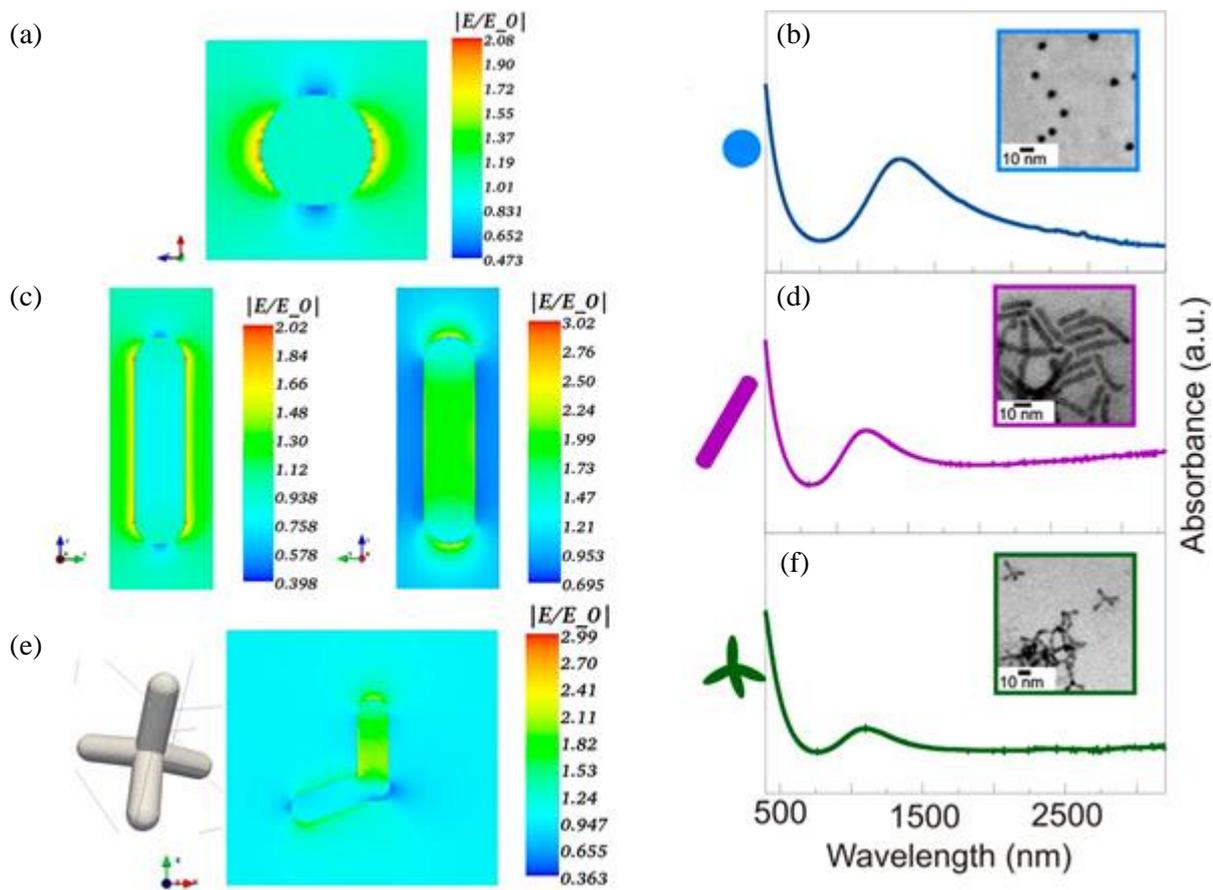

*Figure 17 Simulated hot spots' Near-field enhancement and experimental absorption spectra of nanospheres (a,b), nano-rods (c,d) and tetrapods (e,f). The enhancement of the electric field in the hot spots is limited to a factor of 3 and no low energy band structure is observed in any case. Adapted from* [81]

of spherical symmetry, in fact, generates a multitude of resonant modes as the electric field manifests complex configurations inside the metal inducing regions of high intensity fields, called hot spots. The amplified field in such spots acts as an antenna nearby the nanostructure, increasing field absorption and carrier's generation around the plasmonic material. [82,106,107] The optical properties of metal nanorods, mostly gold and silver, have been extensively studied showing the possibility of manipulating the shape of the plasmonic bands by carefully tuning the aspect ratio (AR) of the nanorods [81,90,106], i.e., the ratio between the long and the short side of the rods. As it can be seen in Figure 15, by modifying the length of the long side of the rod while keeping its width unchanged, a second, more prominent band forms at lower energies while the first band remains almost constant. The high energy peak is located at the same energy of a nanosphere and exhibits a partial blue shift with increasing AR. It is then attributed to transversal modes, charge oscillations taking place along the short side of the nano-rod. [106] The lower energy peak is instead attributed to longitudinal modes.

Stephan Link and Mostafa A. El-Sayed [90] report in their paper a simple, experimental equation for gold nano-rods that shows a linear relationship between the resonance wavelength $\lambda_{max}$ of the low energy band and the AR. Also, the dielectric constant of the medium in which the nano-rods are placed plays a role in the equation. Therefore, the nano-rods can be covered in a semiconductor or oxide shell with a high refractive index. This way it is also possible to extend the transversal mode into the IR and NIR spectra, thus enhancing light scattering and trapping beyond visible light. Similar studies have been conducted on doped semiconductors. I. Kriegel et al. [81] have shown the presence of low energy residuals, rather than well-defined absorption or scattering bands, for three different geometrical morphologies: spheres, rods and tetrapods, all made of $Cu_{2-x}Te$. For the nanorods (AR = 9.6) in Figure 16 (d), the high frequency transversal mode is clearly visible. The blue-shift of the peak with respect to the one exhibited by the nanospheres is in accordance with the previous considerations. The lack of a well-defined energy band for longitudinal modes at lower energies can be attributed to an additional damping term arising at longer wavelengths. [81] Doped semiconductors with disorganized bulk structures exhibit a lower carrier mobility. [108] As a result, the electric fields generated in the hot spots, such as sharp edges and tips, are less intense than they are in noble metals with similar geometries. The increased number of scattering events in doped semiconductors justifies the introduction of an extra damping term for longitudinal modes. In the tetrapod geometry, the electric field enhancement is of the order of 3, much weaker than it is in similar metal geometries. As a result, the absorption spectrum of tetrapods (Figure 16), presents only a weak peak positioned near the one expected for nano-spheres and rods, with no low energy band structures. As scattering events become more frequent, coherence in the charge oscillations is progressively lost. The collective dipole moment then decreases more rapidly, and the chances of scattering events and near-field interactions become negligible. [89] The consequence is a weak shape dependence in non-noble metals plasmonic materials and the suppression of low energy absorption bands.

## 3.5 Hot carriers: generation and transfer

In both the light trapping method and PIRET, plasmonic materials, most often metal nanospheres, are used only as a mean to increase absorption efficiency in the active layer of other structures. Yet, by exploiting Plasmon Induced Charge Separation (PICS), metals or highly doped semiconductors can be used as the active

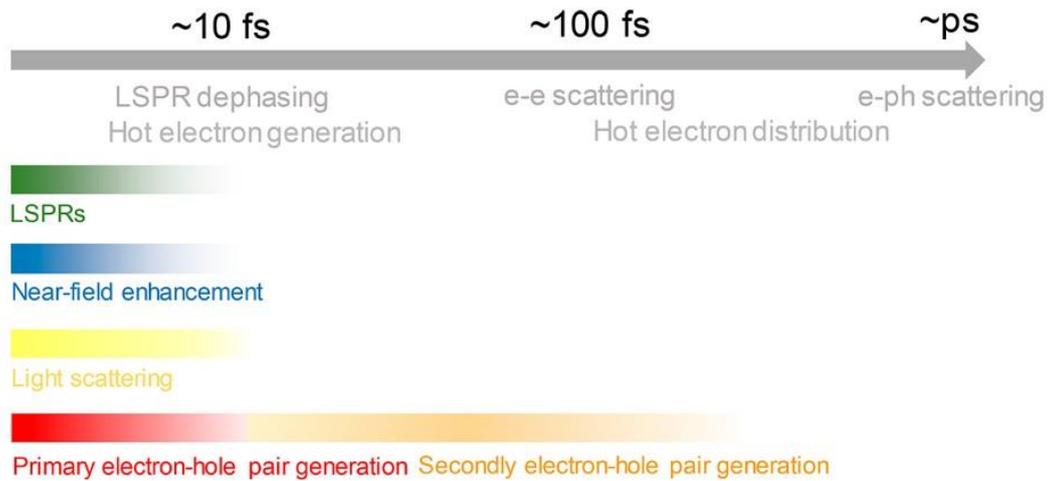

*Figure 18 Time evolution of a plasmonic system. Collective phenomena such as LSPRs, NFE and light scattering can occur only during the first few tens of femtoseconds following excitation, before loss of coherence. e-e scattering and e-phonon scattering cause the quenching of the polarization and the generation of a population of hot electrons. Adapted from [85]*

layer driving the photovoltaic generation in stand-alone applications. This method exploits the energy collected by plasmonic resonances to promote the generation of hole-electron pairs and their consequent injection into the valence and conduction bands of neighbouring layers. The efficiency of each one of the three main processes, hot carriers' generation, injection, and charges recombination determines the overall efficiency of the process. The methods of improvement of hot carriers' generation have been investigated in the previous chapters. Hot electrons are generated when the energy of the radiation, collected by the LSPR, is transferred to the Fermi gas by Landau damping. To increase the number of hot charges it is then necessary to increase the absorption efficiency of the active medium by adjusting size, shape, and spatial distribution of the plasmonic nano-objects. The radiation's energy is initially stored in the nanoparticle as a collective oscillation of conduction electrons whose dipole moment is coupled with the incoming field. The strongly enhanced electric fields generated during these oscillations are responsible for the phenomena of RET and scattering, previously introduced. S. K. Cushing and N. Wu showed in great details the time evolution of the electron population in metal nanospheres following a plasmonic excitation. [89] The dephasing time is defined as the time interval beyond which coherence between the individual electronic oscillators is lost and collective phenomena of field enhancement and scattering become less probable. Figure 17 shows a time scale at which an excited plasmonic system evolves. For metal nano-objects, the dephasing can happen between less than 10*fs* [109] up to 30*fs* [77,89] after the excitement, depending on the spheres' radius. [89] The loss of coherence

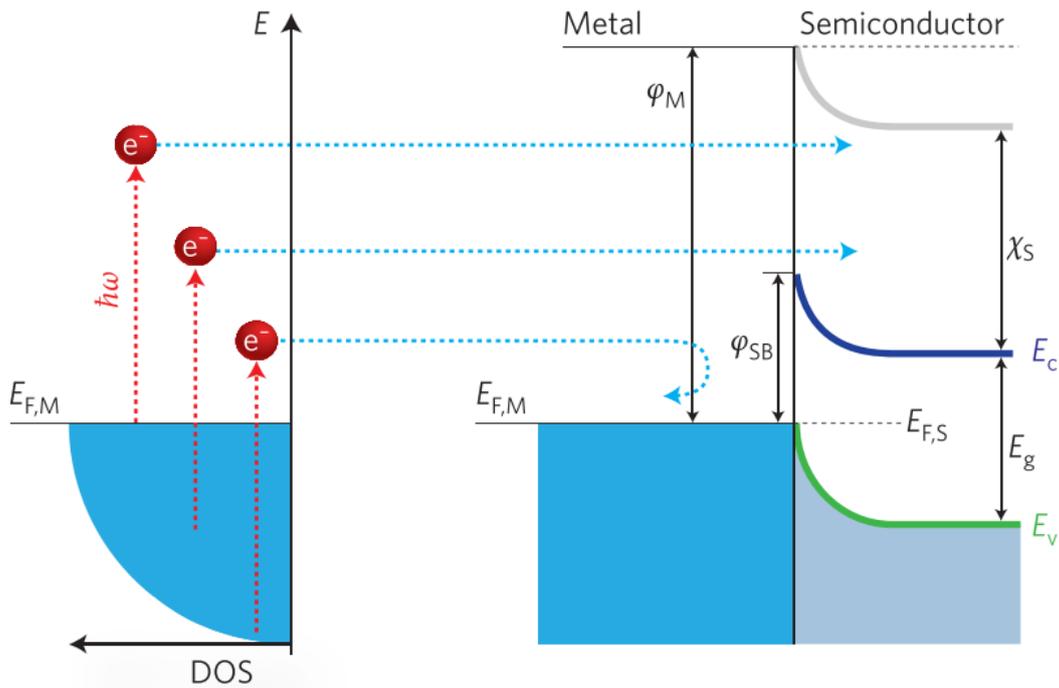

*Figure 19 Hot electrons generated from the proximity of the Fermi Level can have sufficient energy to overcome the Schottky barrier of height $\varphi_{SB}$ formed on the metal/semiconductor interface. In this figure, $\varphi_M$ is the work function of the metal and $\chi_S$ is the electron affinity of the semiconductor. Adapted from [7]*

is a result of Landau damping which transfers energy from the plasmon to the Fermi gas. Elastic scattering events between electrons redistribute such energy generating, in a few hundreds of femtoseconds [85,89,109,110], a Fermi-Dirac distribution of 'hot' carriers [83,85,89,109–111] whose height above the Fermi level is limited by the energy of the photons. [77] Finally, by anelastic electron-phonon scattering the gas cools to equilibrium exchanging its energy with the lattice as heat. If the plasmon is coupled to a semiconductor the hot electrons generated after Landau damping can have sufficient energy to overcome the Schottky barrier (SB) generated on the metal-semiconductor interface and move into its conduction band. The choice of the oxide is essential to achieve efficient PICS as it determines both the height of the Schottky barrier and the rate of injection of hot electrons. [77,78] As electron acceptor, $TiO_2$ is a commonly used semiconductor as its high DOS above the conduction band favours fast injection. [7] The literature is rich of studies conducted for PICS and injection from noble metals such as Au, Ag or Pt, into the conduction band of $TiO_2$ thin films, [112–114] nano-sheets, [115] and nano-tubes. [116] The use of other oxides such as $ZnO$, [117] $CeO_2$ [118] or $SnO_2$ [119] has been reported as well. In doped semiconductors the plasmonic resonance can be found in the IR region at wavelengths longer than $1500 nm$. The height of the barrier can therefore be among the greatest limits to the process efficiency. [7,78] For the electrons to successfully inject into the conduction band of the oxide, the

Schottky barrier needs to be smaller than the photon's energy, yet high enough to prevent electrons from recombining with the holes left in the plasmonic material. The height of the barrier is generally calculated as the difference between the work function $\varphi_M$ of the metal and the electronic affinity $\chi_S$ of the semiconductor (Figure 18):

$$\varphi_{SB} = \varphi_M - \chi_S \tag{3.8}$$

Deeper investigations have shown that, despite the previous equation is in principle correct, the model is incomplete, and it does not account for the non-ideal behaviour of the metal-semiconductor interface. [120] Experimental measurements performed on p-type Silicon-metal interfaces show a weak dependence of the height of the Schottky barrier on the metal work function. Due to the presence of dangling bonds, and therefore a large density of interface states, the Fermi level is pinned at the neutral level causing the SB to be weakly dependent on $\varphi_M$. [120]

## 3.6 Regeneration

To support a stable current flowing from the metal or doped semiconductor into the oxide, the holes left after the electron's injection need to be transported away from the interface as to avoid the rapid deposition of a positive charge. A hole transport material (HTM) or an electron donor material therefore needs to be placed into contact with the active layer to allow charges regeneration. [7,86,121] Also in the case of holes extraction, the correct alignment of the energy bands of the active layer and the HTM is fundamental. When the Fermi energy of the metal, or doped semiconductor, is higher than the ionization energy of the HTM, only hot holes generated after Landau damping can have sufficient energy to overcome the negative energy barrier formed on the active layer-HTM interface. [122] Takuda Ishida et al. [122] presented a work where they deeply analysed the role of different HTM used for hot holes extraction. [122] In their setup, hot electron-hole pairs are generated exploiting LSPRs in gold nanoparticles. Hot electrons are then injected into the valence band of *TiO$_2$* while hot holes are extracted via different HTLs with varying ionization energies. They have shown that a photo-current is measured even when the radiation's energy is lower than the sum of the energy gap between the conduction band of the electron transport material (ETL) and the valence band of the HTL. This result suggests a stepwise injection of hot charges, where only the electron, or the hole, of the photo-generated pair is injected into the neighbouring ETL or HTL while the leftover charge recombines, after relaxation, into the active layer. Seung Hyuk Lee et al. [123] have presented a setup where hot electrons are generated in Indium Tin Oxide (ITO) nanocrystals and injected into the conduction band of $TiO_2$. They have demonstrated that the generated photo-current was in fact optimized by contacting the thin layer of ITO nanoparticles with a thicker layer of $MoO_3$ that acted as an HTL. [123] By choosing a conductor with a proper work function, aligned with the Fermi level of the ITO, it is possible to simultaneously allow holes regeneration while preventing hot electrons from jumping directly into the metal.

## 4. Plasmonic-semiconductor hybrid systems

One of the most significant examples of design and possible applications of hybrid semiconductor nanostructures was reported in the work of Engelbrekt et al. [124], as they discussed the ultrafast photodynamics and structure of 8nm Au-Pt core-shell nanocrystals used for photocatalysis applications. [124] It was reported how by using very small Au nanoparticles encapsulated within an ultrathin layer of Pt, down to just 3 atomic layers, it was possible to combine the broadband light absorption and electric field enhancement properties of the noble metal nanoparticles with the catalytic properties of Pt to realize new highly efficient photocatalytic devices. By means of ultrafast transient absorption spectroscopy and photoelectron spectroscopy it was observed that to maximize the efficiency of these hybrid nanostructures it is important to use Au nanoparticles displaying a very small diameter, as to minimize the scattering of the incoming light and maximize both the absorption and the average local field enhancement introduced by the LSP (Figure 19). Moreover, the small size of the particles also allows to efficiently couple the LSP oscillation with the molecules adsorbed to the surface, maximizing the transfer of electrons from the structure to the molecules and thus strongly enhancing the overall chemical reactivity of the system. Another requirement to be met to optimize the performance of the system is using an ultrathin Pt shell for the structure. In fact, when the outer shell is constituted by a non-plasmonic material with different electron effective mass and carrier density compared to the core, the shell electrons will contribute very little to the LSP. Consequently, a thicker outer layer will only act as an obstacle both to the coupling of the plasmonic oscillation with the surface and to the movement of the hot electrons from the core to the surface, introducing losses and in general lowering

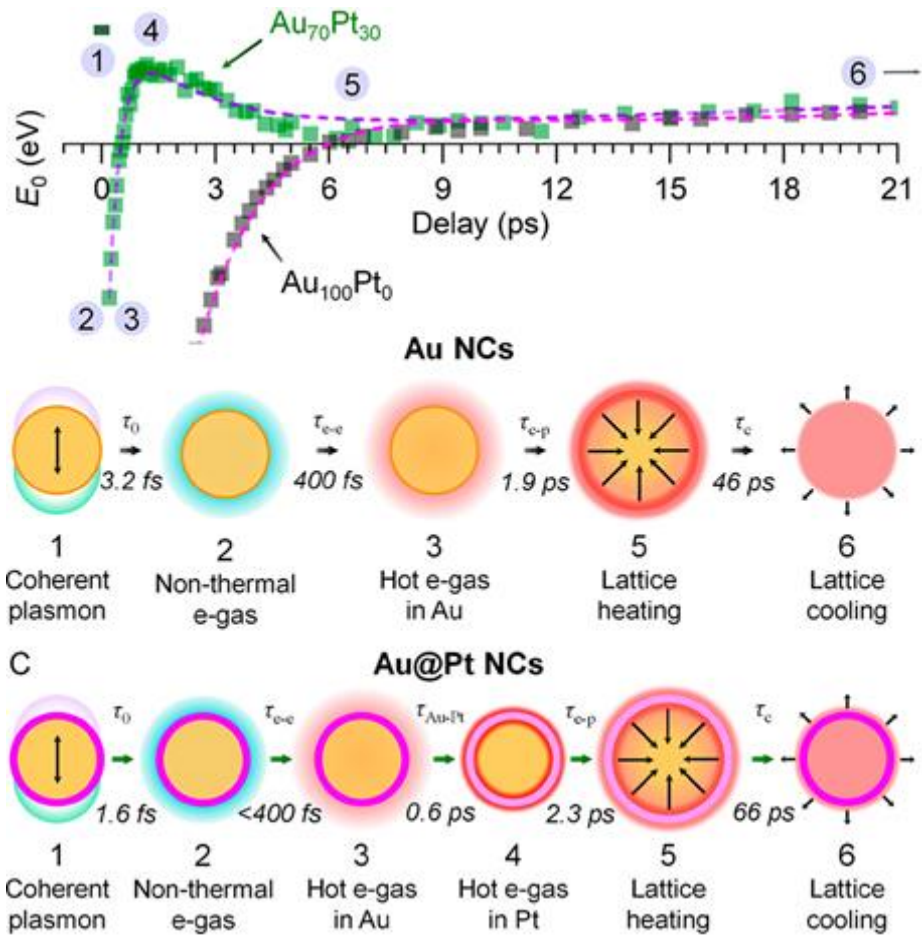

*Figure 20 Main processes and their time scales following LSPR excitation of Au and Au@Pt NCs. (a) Time traces of the LSPR peak energy ($E_0$) for $Au_{100}Pt_0$ NCs (black) and $Au_{70}Pt_{30}$ NCs (green), with the main processes numbered. (b) Schematic sequence and time constants ($\tau$) for the $Au_{100}Pt_0$ NCs. (1 → 2) Plasmon damping followed by creation of a nonthermal electron gas. (2 → 3) Electron thermalization to form a hot carrier gas. (3 → 5) Lattice heating and expansion of the entire NC by electron–phonon equilibration. (5 → 6) NC cooling by heat transfer to the environment. (C) Schematic sequence and time constants ($\tau$) for the $Au_{70}Pt_{30}$ NCs. (1 → 2) Plasmon damping. (2 → 3) Electron thermalization in the Au core. (3 → 4) Electron thermalization on Pt(d) electrons, energizing the Pt surface layer. (4 → 5) Outside-in lattice heating. (5 → 6) NC cooling. Reprinted from* [124]

both the number and the energy of the hot electrons transferring to the outer surface. The choice of the material used for the shell is not only important to maximize the photocatalytic efficiency, but also to extend the overall light absorption range of the structure by combining high values of the absorption coefficient with the plasmonic field enhancement. Also acting on the shape of the hybrid nanostructure is important to achieve higher efficiencies, as the shape and size of the system influence both the scattering/absorption properties and the coupling of the LSP to the adsorbate. Several shapes have been investigated by various groups, including core-shell nanospheres, nanorods and triangular nanoprisms. [124–126]

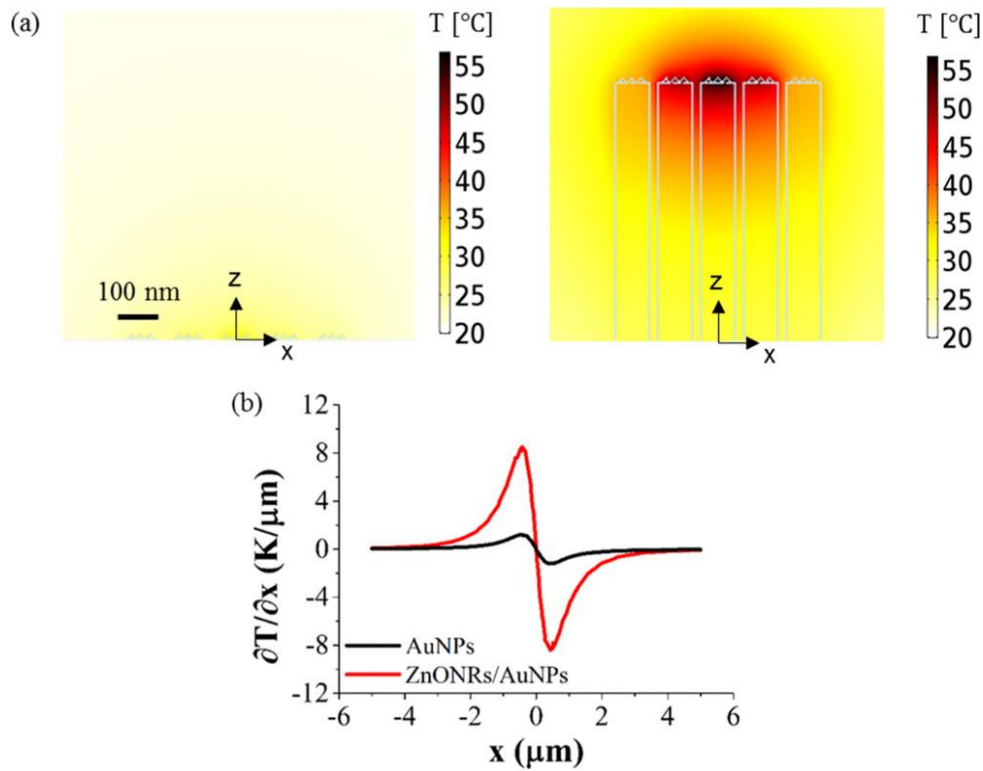

*Figure 21 Comparison of the plasmonic heating effect between the AuNPs and the ZnONRs/AuNPs. (a) Simulated temperature fields around the AuNPs (left) and ZnONRs/AuNPs (right) at y = 0. (b) Simulated temperature gradient along the x-direction for the AuNPs (black line) and ZnONRs/AuNPs (red line). Reprinted from [127]*

A similar work was also performed by Li et al. [128], who investigated the possibility to realize 3D hybrid nanostructures exploiting both noble metal nanoparticles and monolayered materials, more specifically Au, Ag and $MoS_2$. [128] The structures that were investigated were constituted by a large Au-$MoS_2$ core-shell nanoparticle to which smaller Au/Ag nanoparticles were attached, so that the final system would display an extremely high density of plasmonic hot spots for the electric field due to the $MoS_2$ layer acting as a spacer between the different plasmonic nanoparticle spheres. The results obtained were different with respect to previous similar works where graphene was used as a spacer. The main difference was the absence of the quantum mechanical electron tunnelling effect, which was observed in the case of graphene displaying just a 3.4 Å thickness for a single layer. The electronic tunnelling leads to an equilibration of the charges inside the nanostructure, thus lowering the plasmonic field enhancement and also the number of hot spots. $MoS_2$ proves therefore to be a better choice to realize this kind of hybrid system, allowing to achieve extremely high field enhancement, also thanks to its capability of exchanging electrons with the Au and Ag nanoparticles. This kind of nanostructure is therefore extremely promising for photocatalysis, surface enhanced Raman scattering (SERS) and sensing applications.

Another interesting application of hybrid plasmonic systems is the possibility to realize very efficient plasmonic optical tweezers. These could be useful in fields like biology and chemistry to achieve light trapping of small objects like nanoparticles, molecules, and bacteria. In a paper from Lee et al. [127] from 2018 it was reported that by coupling gold nanoparticles to zinc oxide nanorods a noticeable improvement of plasmonic heating and near-field enhancement was achieved, allowing to realize optical tweezers that exploit plasmonic effects to reduce the laser power needed to operate (Figure 20). [127] This is important as smaller objects experience lower optical gradient forces, and therefore require higher beam intensities in order to be properly trapped. This can easily lead to instabilities in the system and to the possibility of damaging the trapped object, especially in the case of bacteria and cells. Instead, by using plasmonic effects it is possible to achieve near-field enhancement and heating to obtain confinement with lower laser powers and avoid the aforementioned problems. Several studies had been conducted to investigate different possible shapes for the gold nanostructures used for the tweezers, including nanopyramids, nano-islands and nano-bowties, but remarkable results have also been obtained by employing hybrid, more complex structures displaying synergistic effect between different materials. In particular, in the work by Lee et al. [127] the increase in the performance of the nanostructure was attributed to the capability of the ZnO nanorods to strongly absorb the incoming light and transfer its energy to the gold nanoparticles through leaky waveguide modes. This effect allowed to increase the overall efficiency of the hybrid system, storing more energy in the plasmonic mode and leading to a stronger heating. It is crucial in this application as the medium natural convection and the thermophoretic force are the two main factors contributing to the particle trapping. A third possible application for hybrid plasmonic nanostructures involves molecules sensing and DNA sequencing. This can be useful when considering systems displaying nanopores like those discussed by Garoli et al. [129], where $MoS_2$ flakes were integrated in a structure characterized by Au nanoholes. [129] In this case, the aim is to couple the plasmonic field enhancement from the nanocrystals to the in-plane electric field localization that characterizes 2D materials like $MoS_2$, graphene, etc. to achieve a new generation of high efficiency, high throughput devices for molecule sensing (Figure 21).

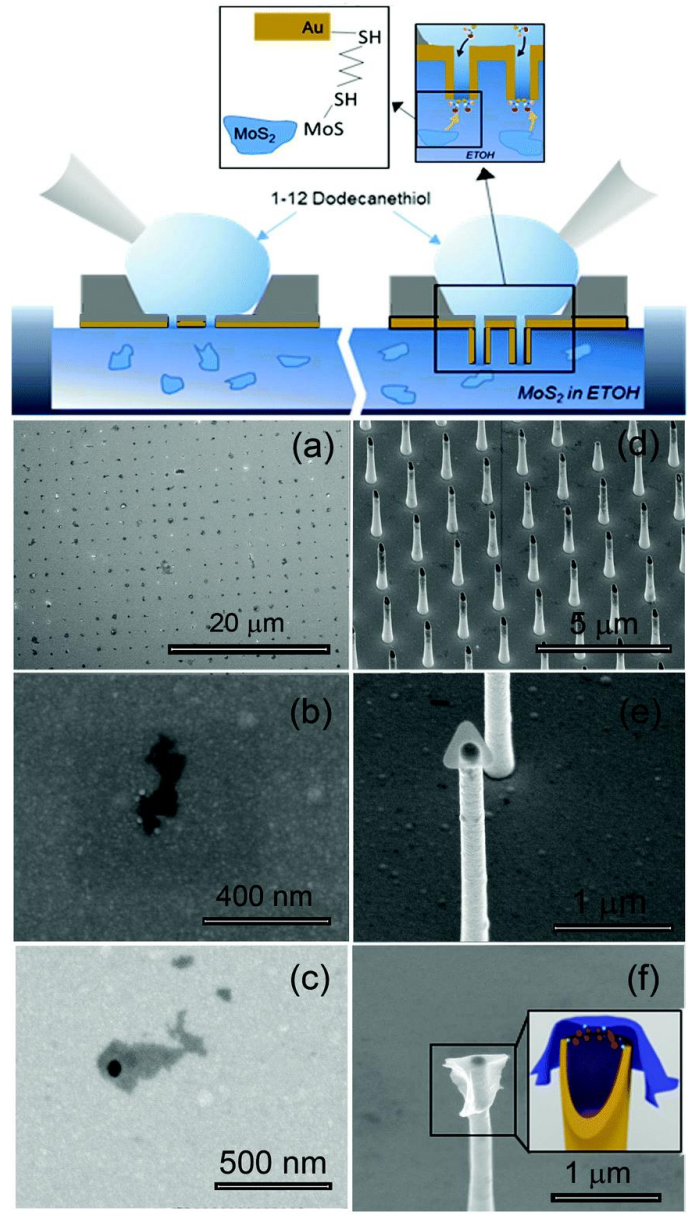

*Figure 22 SEM micrographs of MoS$_2$ flakes deposited onto an array of plasmonic nanoholes. (Top panel) Illustration of the concept for controlled deposition of MoS$_2$ flakes over metallic holes; (a) top view over a large flat gold hole array; (b) and (c) details of a single-layer flake on a 2D pore; (d) tilted view over a large array of 3D antennas covered with MoS$_2$ flakes; (e) and (f) details of the MoS$_2$ flakes deposited onto an antenna. Reprinted from* [129]

In this kind of devices, the fluorescence of molecules traversing the nanostructure is strongly increased thanks to the enhancement introduced by the nanopore. For this reason, 2D materials represent the ideal solution to realize solid state membranes to be used for sensing, but they need to be attached to a larger structure displaying nanoholes as they are atomically thin. In this kind of technology, the challenge resides in finding a stable, reproducible, and scalable way of realizing the membrane, which can be realized either through high energy transmission electron microscope milling or by focused ion beam milling. In the work reported by Garoli et al. [129] it was demonstrated that focused ion beam milling is a safer and more reproducible way to realize nanoholes, and that the defects of the monolayered material allow to have nucleation of additional nanoparticles to further enhance the sensitivity of the system.

# 5. Optical methods of investigation

5.1 Ultrafast transient absorption spectroscopy

One of the critical matters to be addressed when studying hot-electron-transfer based devices is properly understanding and reconstructing the real dynamics of the hot carriers that are generated inside the material after the excitation from the impinging photons. A first step to be taken to address this issue is performing mathematical modelling and simulating the nanostructures that are going to be implemented. This is commonly done by using finite elements methods (FEM) and performing density functional theory (DFT) simulations. [130–132] This way it is possible to identify the best shapes and dimensions for the nanostructure that one wants to employ, thus maximizing the desired effects, while also retrieving the expected distribution of the electric field and of the excited carriers. This also allows to make assumptions on the behaviour of the real structure that will be later fabricated and investigated. This can be done both for monometallic plasmonic nanostructures and hybrid systems, especially for noble-metal-based devices, although the computational cost of the simulation can be high the more complex the structure becomes, especially in the case of DFT. In fact, the jellium model is typically employed to simplify the quantum mechanical description of the system, and it performs very well when simulating monometallic structures. Unfortunately, this model cannot be applied in the case of hybrid nanostructure, as it does not allow to describe properly the interfacial effects and electronic states which play a major role in the hot electron generation and transfer processes. [133] Moreover, for more complex plasmonic materials like doped semiconductor nanocrystals (e.g. ITO, FICO, AZO, $CuSe_{2-x}$) the possibilities to perform accurate quantum mechanical simulations are limited, as the processes involving hot electrons are strongly influenced by material defects which cannot be easily modelled at computational level. In both cases, whether the aim is to confirm the prediction of the theoretical simulation or if there is no possibility to have a very accurate prediction and the sample must be investigated directly, an accurate time

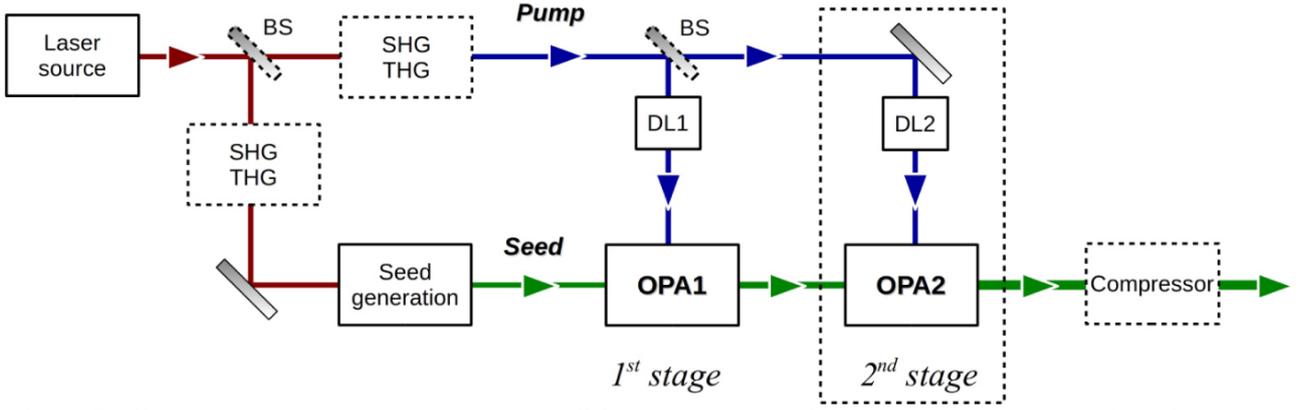

*Figure 23 Conceptual scheme of a femtosecond OPA: the beam coming from the laser source is split and used to generate both the pump of the OPA, which can be either the fundamental or its second or third harmonic, and the seed, achieved by a white light generation process. The OPA is then realized by achieving spatial and temporal overlap of the two beams on a suitable nonlinear crystal. It is then eventually possible to achieve further amplification or generation of longer wavelengths by difference frequency generation through an additional stage. DL: delay line; BS: beam splitter; SHG/THG: second/third-harmonic generation module. Dashed boxes denote optional stages. Adapted from* [134]

resolved measurement must be performed to correctly reconstruct the hot electrons dynamics. The best solution to do this, given the extremely fast temporal dynamics of the processes taking place inside the material, is to perform ultrafast transient absorption measurements on the plasmonic nanostructures. [135,136]

In this experimental technique two different pulses are sent on the same spot of the sample under investigation: a first one, called pump, triggers the process that needs to be reconstructed, while the second one, called probe, is delayed by a time $\tau$ with respect to the pump and it is used to determine which physical processes are occurring inside the material. Following the interaction with the pulses, a signal which is proportional to the third order polarization is emitted along the same direction of propagation of the probe, which experiences a phenomenon called self-heterodyning. [137] The transmitted intensity of the probe $I(\lambda, \tau)$ is detected as a function of the probe's wavelength $\lambda$ and the reciprocal delay $\tau$ between the pulses. This signal is proportional to:

$$I(\lambda, \tau) = \left|\tilde{E}_{probe}(\lambda) + \tilde{E}^{(3)}(\lambda, \tau)\right|^2 \cong \left|\tilde{E}_{probe}(\lambda)\right|^2 + 2\left|\tilde{E}_{probe}(\lambda)\right| Im\{\tilde{P}^{(3)}(\lambda, \tau)\} \qquad (5.1)$$

where $\tilde{E}_{probe}(\lambda)$ is the electric field of the transmitted probe for the unperturbed sample and $\tilde{E}^{(3)}(\lambda, \tau) \propto i\tilde{P}^{(3)}(\lambda, \tau)$ is the signal arising from the sample. It is proportional to the third order polarization of the material $\tilde{P}^{(3)}(\lambda, \tau)$ which contains information on the lifetimes of the excited electronic states of the material.

Therefore, by comparing the transmission of the perturbed and unperturbed sample it is possible to retrieve the differential transmission $\Delta T/T$ of the probe, which is a function of both the probe wavelength and the delay $\tau$. [137,138] The differential transmission is defined as:

$$\frac{\Delta T}{T}(\lambda,\tau) = \frac{T_{on}(\lambda,\tau) - T_{off}(\lambda)}{T_{off}(\lambda)} = \frac{2Im\{\tilde{P}^{(3)}(\lambda,\tau)\}}{|\tilde{E}_{probe}(\lambda)|} \qquad (5.2)$$

where $T_{on}$ is the transmission of the probe when the pump is exciting the sample, while $T_{off}$ is the probe transmission when the sample is unperturbed. This quantity can be easily retrieved by modulating the pump beam at a frequency which is half of the repetition rate of the laser feeding the system, so that half of the pump pulses are blocked, and half are transmitted. The differential transmission can have both positive and negative values, the former being associated either to ground state bleaching (GSB) or stimulated emission (SE), while the latter is associated to excited state absorption (ESA). [139] By varying the pump-probe delay $\tau$ it is possible to retrieve the temporal dynamics of the processes, with the available time resolution being limited from duration of the convolution between the instrumental response function of the system and the temporal profile of the pump pulses. [140] Therefore, in order to perform an ultrafast transient absorption experiment the most important conditions to be met are: 1) use of ultrashort pump pulses to excite the system, which can be achieved in many different spectral ranges by introducing proper compression of pulses generated, for example, from non-collinear optical parametric amplifiers (NOPA) (Figure 22); [141–143] 2) use of a broadband probe in order to maximize the information content obtained from a single shot measurement, thus also improving the overall measurement speed; [140] 3) tuning of the excitation pulses to be resonant with the system under investigation, in order to trigger the process that needs to be studied. With this experimental technique it is thus possible to reconstruct ultrafast dynamics ranging from the ultraviolet to the near infrared.

An example of application of this technique for the investigation of plasmonic nanostructures has been reported in a work from Wu et al. [144], where ultrafast transient absorption measurements were performed on a system composed of gold nanospheres attached to CdSe nanorods. [144] To investigate the possibility of achieving plasmon induced charge transfer, where the decay of the plasmonic resonance directly excites an electron from Au to the CdSe conduction band, the samples were excited at 800nm, in resonance with a plasmonic mode damped by the coupling of the nanosphere with the electron acceptor. After the excitation, both the 1Σ exciton bleaching at 580nm and intraband absorption at 3000nm were observed, confirming the transfer of hot

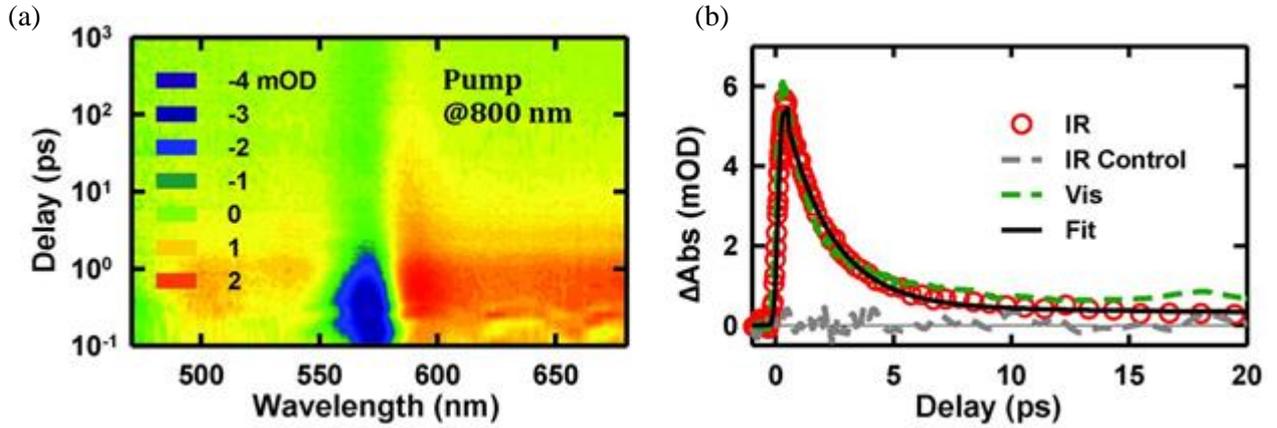

*Figure 24 Plasmon-induced charge separation in CdSe-Au NRs: (a) Two-dimensional plot of transient absorption spectra for CdSe-Au NRs at 800-nm excitation. (b) Intraband absorption (probed at ~3000 nm, red circles) and 1Σ-exciton-bleach (~580 nm, green dashed line) kinetics of CdSe-Au NRs after 800-nm excitation. A negligible intraband absorption signal is apparent in a control sample of a mixture of CdSe NRs and Au nanoparticles (gray dashed line). The black solid line is a multiexponential fit of the kinetics. Reprinted from [144]*

from the Au nanoparticle to the conduction band of the CdSe nanorods (Figure 23). The same signal could in fact be observed for CdSe nanorods/CdSe nanorods-benzoquinone systems when pumping at 400nm, above the CdSe bandgap. The pump used in the measurement for the Au-CdSe nanostructure was instead below the CdSe bandgap and the photons cannot directly excite the CdSe. On the other hand, it was possible to promote electrons from the Au nanoparticles. The matching of the temporal dynamics observed at 580nm and 3000nm with those obtained for the CdSe systems without Au also provide another proof of the proposed electron transfer mechanism, as the same dynamics are absent for two separate solutions of CdSe nanorods and Au nanoparticles. By fitting the measured temporal dynamics, it was possible to retrieve the time constants for the hot electron transfer and charge recombination processes, while by performing polarization dependent measurements it was possible to determine the anisotropy of the sample. [144] This could be done by measuring the transient absorption signal with both parallelly and perpendicularly polarized pump and probe, and then calculating the anisotropy factor $r$, defined as:

$$r = \frac{S_\parallel - S_\perp}{S_\parallel + 2S_\perp} \quad (5.3)$$

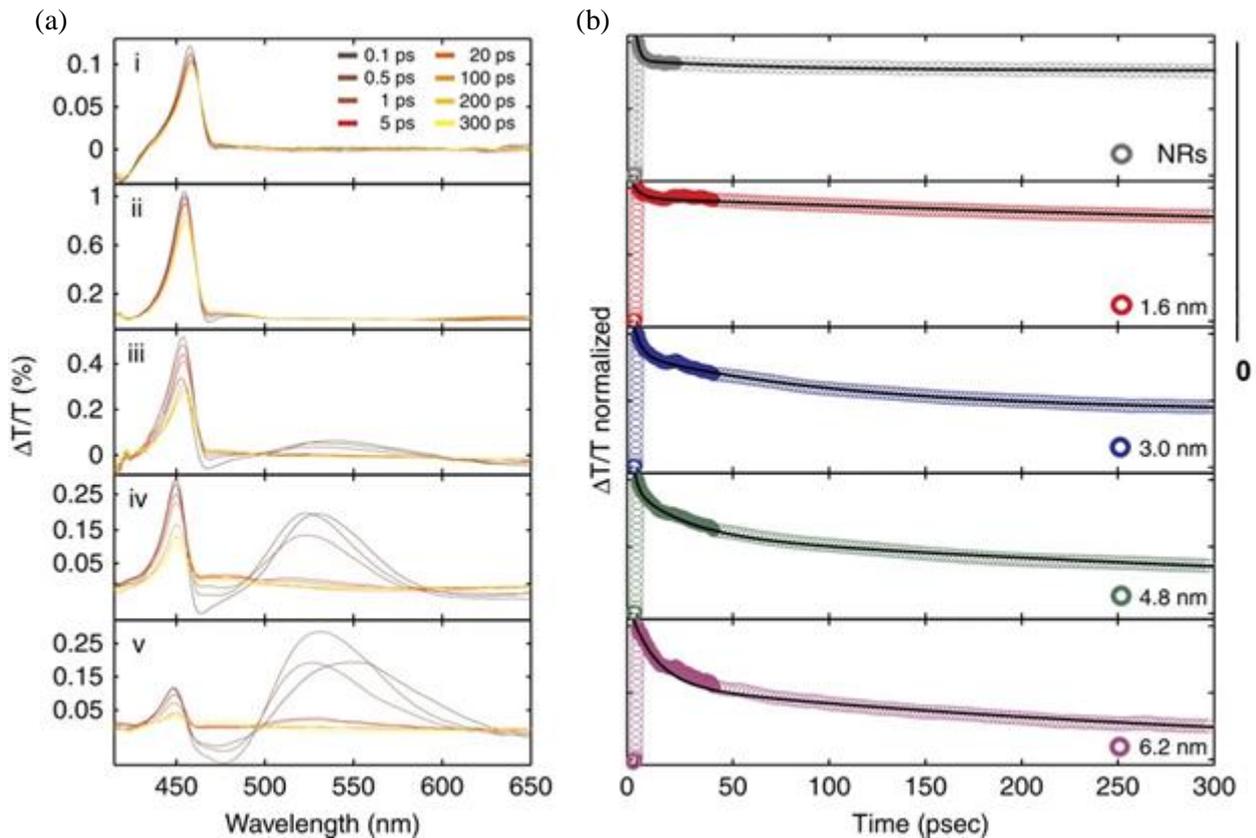

*Figure 25 (a) Transient absorption spectra of CdS NRs (i) and CdS-Au hybrid nanoparticles for different Au metal tip sizes including 1.6 nm (ii), 3.0 nm (iii), 4.8 nm (iv) and 6.2 nm (v) at 450 nm excitation. (b) Corresponding normalized transient absorption dynamics of the bleach recovery at 450 nm, attributed to the first excitonic transition of the CdS NR component for CdS NRs and CdS-Au hybrid nanoparticles with different Au metal tip sizes (reported in each plot). Reprinted from [136]*

where $S_\parallel$ is the signal measured for parallel polarizations while $S_\perp$ is the one measured for perpendicular polarization. Such measurement is particularly useful for plasmonic nanostructures as plasmonic modes are usually polarization sensitive. As a matter of fact, some nanostructures display preferential directions for the oscillation of the electrons, so that one polarization couples more energy into the plasmonic mode with respect to the other. [145–147]

Other examples of the use of ultrafast spectroscopy to investigate plasmonic nanostructures were reported by Ben-Shahar et al [136]. for systems of cadmium sulfide-gold nanorods, to determine the effect of the Au nanoparticle size on the photocatalytic properties of the structure and charge transfer dynamics (Figure 24). [136] In another work by Engelbrekt et al. [124] Au-Pt core-shell hybrid heterostructures were investigated through this technique, [124] while Catone et al. [148] reported the results of ultrafast spectroscopy measurements performed on several systems, including ZnSe and Si nanowires, $CeO_2$-Au NP, ZnSe-Au NP and ZnSe-Ag NP hybrid structures. [148] Finally, Ran et al. [149] retrieved the single exciton lifetime and electron transfer time for type II CdZnSe/ZnSe core-shell quantum dots. [149]

5.2 Two-photon photoemission spectroscopy

Another very useful technique which can be used to investigate the properties of plasmonic systems is represented by two-photon photoemission spectroscopy (2PPE). [150] This technique is similar to ultrafast transient absorption but complementary in terms of the information that it allows to retrieve. In this experiment the sample under investigation is excited by using two different pulses, a pump and a probe, so that the electrons of the material are promoted to a state above the vacuum energy after the excitation by both pulses, thus achieving photoemission. The energies of the photons of the individual pulses are chosen so that they are too low to directly excite an electron outside of the material. For the process to take place the following relations must be met:

$$h\nu_{pump} < \phi , h\nu_{pump} + h\nu_{probe} > \phi \qquad (5.4)$$

where $\nu_{pump}$ and $\nu_{probe}$ are the frequencies of the pump and the probe respectively, while $\phi$ is the work function of the material. Two possible excitation pathways are possible for the electrons: a first one where the electrons are initially promoted from a surface state to an image potential state by the pump, and then to a vacuum state by the probe; a second one where a coherent two-photon excitation directly promotes the electron to the vacuum state. [150] By measuring the angle and the kinetic energy of the photoemitted electrons it is possible to retrieve a 2PPE spectrum, which yields information on the band structure of the material by identifying peaks associated to both surface and image potential states. It is also possible to extend this technique to a time resolved approach by varying the delay between the pump and probe pulses, thus reconstructing the temporal dynamics of the excitation and relaxation process, retrieving the lifetimes of the electronic states (Figure 25). [151,152] This experimental technique is particularly useful to study surfaces, adsorbate-metal systems and plasmonic nanostructures. In a recent work, this technique was employed by Foerster et al. [153] to verify the hypothesis that plasmons decay into the same interfacial states of hot electrons for various metal-oxide interfaces, including Au-HfO$_2$ and Au-TiO$_2$ thin films. [153] Two-photon photoemission spectroscopy was used to retrieve the lifetimes of excited electrons, which were then compared

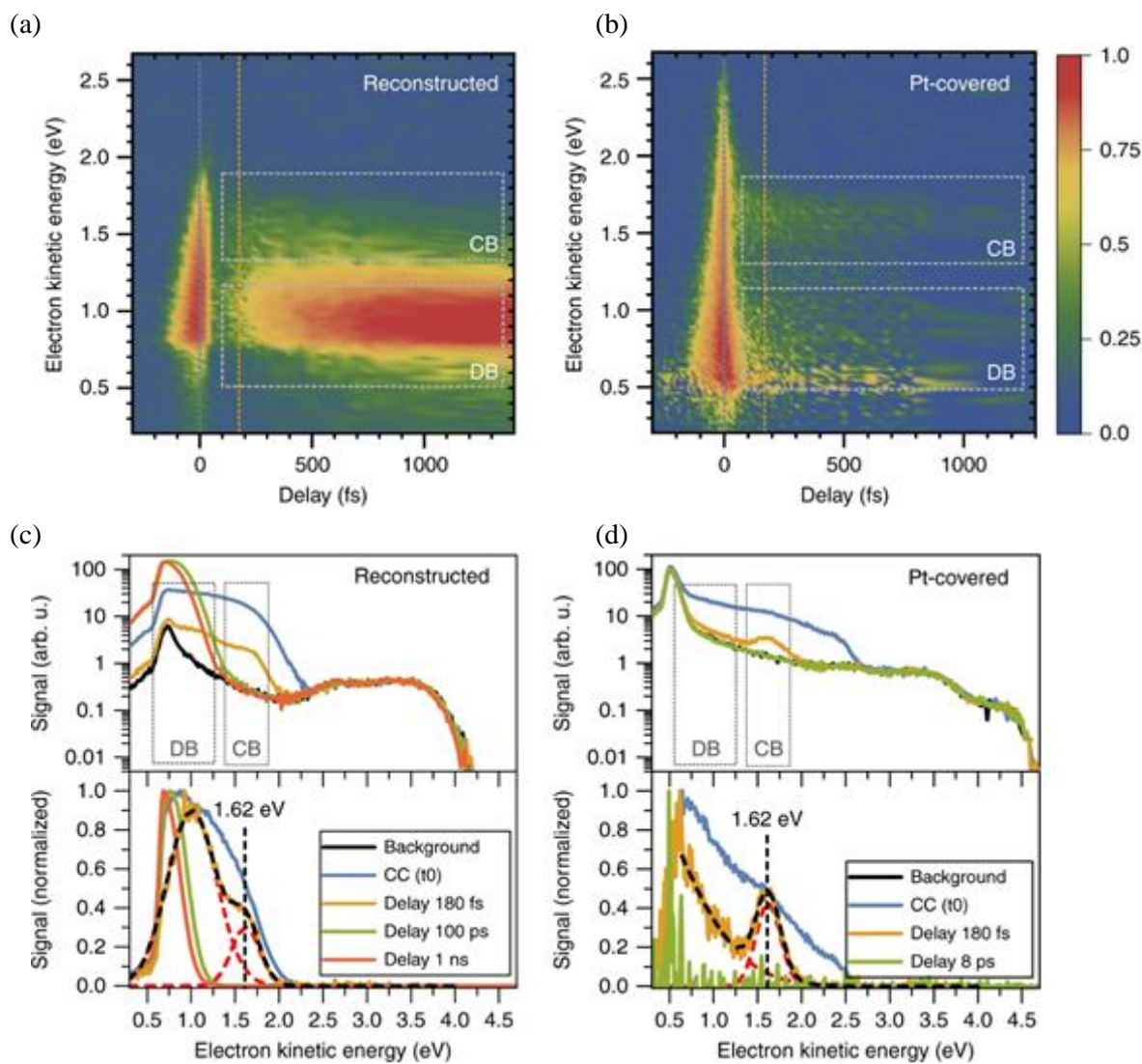

*Figure 26 An example of TR-2PPE spectra. Color maps of the transient photoemission signal as a function of electron kinetic energy and pump-probe delay for reconstructed Cu$_2$O (100) surface (**a**), and for Pt-covered Cu$_2$O (**c**). In **b** and **d**, selected spectra are shown at the specified pump-probe delays before (top) and after (bottom) background subtraction. In **a**–**d** the positions of the conduction and defect bands are indicated (CC: cross-correlation, DB: defect band, CB: conduction band). Reprinted from* [151]

with the ones obtained from the resonance line width observed in single particle scattering spectra, obtaining good quantitative agreement for the damping of hot electrons and plasmons. Other examples of use of 2PPE spectroscopy to characterize plasmonic systems include a study by Tan et al. [109], where coherence and hot electron dynamics in Ag nanocluster/TiO$_2$ heterojunctions were investigated. [109] 2PPE microscopy was also used by Frank et al. [154] to study the dynamics of short-range surface plasmon polaritons in single-crystalline gold platelets deposited on a SiO$_2$ substrate. [154]

## 5.3 Two-photon absorption

A third experimental technique that can be used to study plasmonic systems, especially to investigate sub-diffraction regions, is represented by two-photon absorption (2PA) measurements. [155] In this experiment the sample to be inspected is illuminated with a tightly focused beam constituted by photons having energies lower than the material's bandgap for semiconductors, and around half of the plasmonic resonance energy for metal systems. The experiment can be performed in two different ways. It is possible to realize a direct measurement, where the excitation beam is focused in a point and the sample is scanned in and out of the focus through a Z-scan translation stage, while the transmitted portion of the beam is measured with a spectrometer. This way, the measured intensity will have a minimum when the sample is in the focus, provided that it displays two-photon absorption at that specific wavelength. By fitting the Z-scan data it is then possible to retrieve the two-photon absorption cross-section for different wavelengths, reconstructing the nonlinear excitation spectrum of the sample. [149,155] The advantages of this approach are the simplicity of the experimental setup and the possibility to realize the experiment with any sample. The main drawback is that the detection takes place in a non-background-free direction, so that high sensitivities are required to measure the difference in absorption. Another possibility is to perform an indirect measurement, exciting the sample and detecting the supercontinuum emission generated by its relaxation to the ground state. This way it is possible to correlate the measured spectrally integrated intensity with the efficiency of the two-photon absorption process, as higher emitted intensities will be associated to higher 2PA cross-sections. [155,156] The advantage of this approach is that the emission takes place in a background-free direction, so that it is easy to measure the signal, while the drawback is that not all samples may display photoluminescence after the excitation.

Examples of application of this technique include a work by Borys et al. [156], where the indirect approach was used to investigate the delocalization of plasmonic modes in silver nanoparticle aggregates, revealing it through the excitation of nonlinear hot spots. [156] Samples displaying different thicknesses for the Ag layer were inspected to study how the plasmonic modes are affected by the coupling between the individual nanoparticles. The samples were realized by Tollens reaction using different growth times, so that they ranged from well separated nanoparticles for short growth times to progressively more continuous nanostructures for longer growth times. The spectral characteristics of the plasmonic modes were reconstructed by exciting the sample in the IR over a wide range of energies and indirectly measuring two-photon absorption, whose efficiency is enhanced due to the plasmonic confinement of the electric field in sub-diffraction regions of the sample called 'nonlinear hot spots' (Figure 26). It was observed that the spectral width of the resonances is broader for the individual nanoparticles and becomes progressively narrower for longer growth times, with the semicontinuous Ag layer displaying multiple narrow resonance peaks. This happens because as the contact between the different nanoparticles increases the LSPRs of the individual nanoparticles become progressively hybridized with the delocalized plasmonic modes typical of larger nanostructures. This is also confirmed by the fact that the location of the supercontinuum emission for the semicontinuous Ag layer does not always correspond to the excitation site, differently from the case of individual nanoparticles, where the two coincide. The progressive delocalization of the plasmonic modes was also confirmed by polarization anisotropy measurements. These highlighted the greater sensitivity of individual nanoparticles to s-polarized light, diffractively coupled into the modes. On the other hand, the semicontinuous layer shows greater sensitivity to p-polarized light. The increased absorption for p-polarized light in denser structures is similar to what is observed for propagating surface plasmons in smooth metal films, which need a component of the electric field perpendicular to the surface in order to be excited.

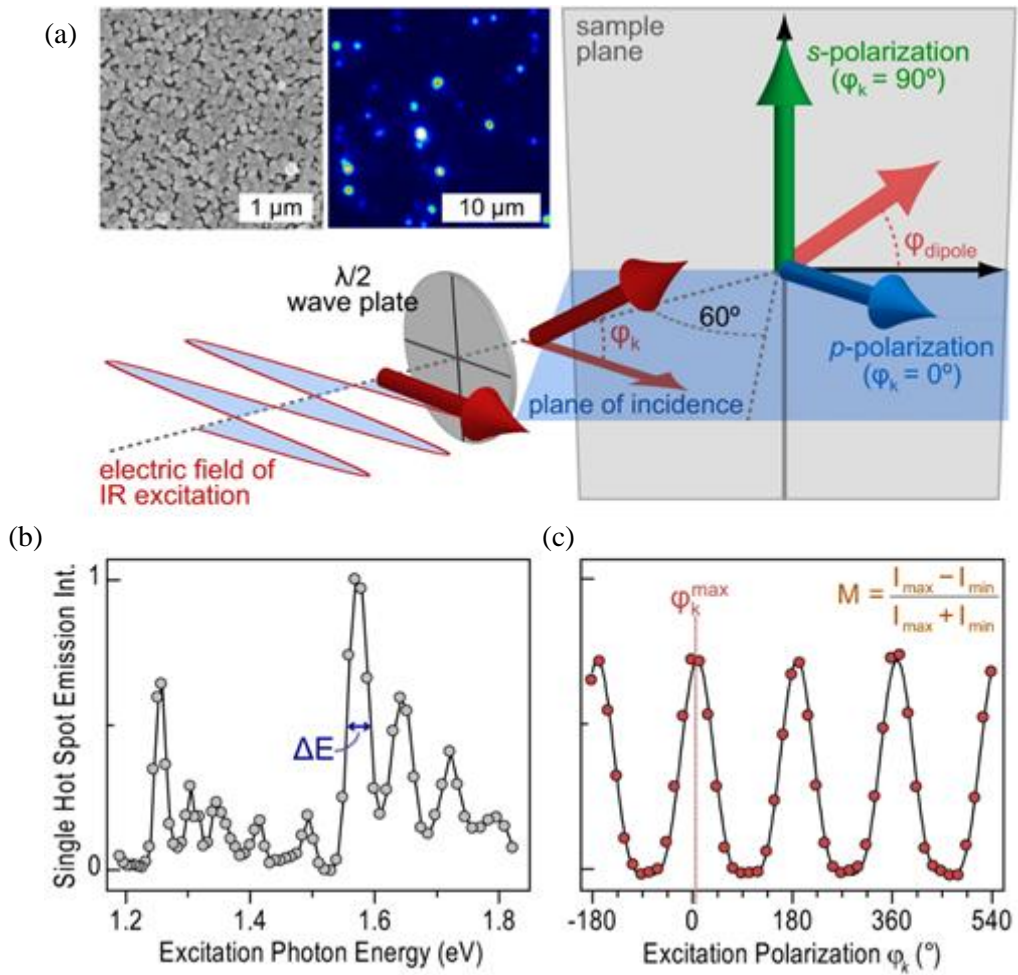

*Figure 27 Excitation spectroscopy and polarization anisotropy of single nonlinear hot spots in a high-coverage, semicontinuous silver film grown following the Tollens mirror reaction: (a) under illumination of obliquely incident, pulsed infrared radiation, the film exhibits surface-enhanced upconverted supercontinuum emission from discrete, diffraction-limited spots. Typical SEM and fluorescence micrographs are shown; (b) excitation spectrum of a single hot spot where the emission intensity is plotted as a function of the IR photon energy. The hot spot exhibits multiple resonances, each characterized by their own FWHM, ΔE; (c) excitation polarization anisotropy measurements on single hot spots. The emission intensity is recorded as a function of the angle $\phi_k$ of the electric field vector from the horizontal defined in the cartoon and can be fitted with a $cos^4$ function (black line). The coupling to polarized excitation is characterized by the polarization anisotropy value M, defined in the inset and the in-plane dipole angle, $\varphi_{dipole}$, which is determined from the angle of maximum emission $\phi_k^{max}$. Reprinted from [156]*

Other examples of application of this technique to investigate the optical properties of plasmonic nanocrystals include a work from Marin et al. [157], where the 2PA and two-photon-induced emission (2PE) properties of nanodisks of $Cu_{2-x}S$ in the covellite phase were studied. [157] The final aim of this research is to exploit inorganic doped semiconductor nanoparticles, which display light upconversion capabilities in the NIR spectral range, for multiple different biophotonic applications, like photodynamic therapy, clinical diagnostics, and fluorescence imaging of living tissues. The advantages of using such materials stand in the possibility of increasing the 2PA efficiency by exploiting the plasmonic enhancement, guaranteed by their high carrier densities, and displaying absorption in the NIR spectral range. This last characteristic is important as it allows to use wavelengths in this spectral region to probe living tissues, achieving a longer penetration depth, with respect to visible light, thanks to reduced scattering events. Moreover, said inorganic nanoparticles display lower photobleaching and higher photostability compared to conventional organic dyes.

While the 2PA properties of plasmonic nanoparticles have more direct applications in the field of biophotonics, especially when using them as markers for the imaging of living tissues, investigating the 2PA cross section can in general be useful to fully reconstruct the plasmonic properties of this kind of materials. One possible application is the extension of the experiment reported in the work by Borys et al. [156] to study the resonances of doped semiconductor nanocrystals, usually characterized by extremely broad plasmonic resonances in the NIR spectral range, between 1-2 μm. When performing far field absorption measurements, a spatial average of different plasmonic modes is usually inspected. Instead, by performing 2PA experiment it is possible to excite much smaller regions of the samples. This way, one can determine whether the wide plasmonic peaks usually measured are given by a superposition of many peaks, associated to different modes. By knowing the true width of the plasmonic resonance it would be possible to determine the actual relaxation time of the excited electrons, useful to reconstruct the real dynamics of the electron-electron scattering processes. Since the resonances displayed by doped semiconductor nanocrystals lie in the NIR spectral range, performing two-photon absorption experiments would require an excitation source emitting in the mid-IR range. This could be achieved by using appropriate optical parametric amplifiers in the region of interest. [141]

All these optical spectroscopy techniques are extremely useful to study hot electron extraction from doped semiconductor-based heterojunctions. Deep understanding of the physical processes is necessary for the advancement of the field to reach efficient harvesting of infrared radiation and increased overall efficiencies.

# 6. UV Harvesting

Finally, we would like to mention another strategy to broaden the Sun's irradiation harvesting by solar devices. The region of the sun spectrum identified by wavelengths below 400nm has generally been neglected by photovoltaic technologies as most of it is filtered by Earth's atmosphere. Even if there is less overall photon flux in the UV, theoretical calculations predict an achievable efficiency of 7%. [158] Moreover, harnessing those photons with higher energy and achieving full spectral response could represent the right way to increase power conversion efficiency of solar cells. Many studies showed UV light to be deleterious for organic and perovskite-based photovoltaic devices leading to degradation of the active material or other components of the cells and reduced performances over time. [25,159] Most of the times some filters are used to shield the devices from UV light, preventing them from damage but also increasing costs. Being able to harvest UV light before it reaches UV-sensitive layers could mean giving a boost in efficiency and avoid the use of filters, in turn leading to a simultaneous lower cost and increased efficiency. In 2018, Hee-Suk Roh et al. [160] successfully demonstrated the use of Phosphor-in-glass (PiG), namely phosphor particles in a glass matrix, as a down-converter for a UV-stable perovskite solar cell. [160] Such devices showed increased photon conversion efficiency at 320 nm and resistance to UV degradation. [160] In the same year, D. Liu et al. fabricated different UV-harvesting perovskite films by spin coating, showing the tunability and versatility of halide perovskite materials. [158]

# 7. Conclusion

In this review we have discussed the possibility to harvest the infrared part of the solar irradiation at the ground level with different materials and systems, such as multijunction solar cells, inorganic-organic perovskite solar cells, organic solar cells, quantum dot solar cells and upconverting materials based solar cells. We have focused our attention on plasmon induced hot electron extraction, which occurs in a heterojunction between a plasmonic material and a semiconductor. If the plasmonic material is a doped semiconductor, its absorption is in the infrared. Moreover, by varying dopant concentrations and geometries, the position of the absorption bands can be properly modulated to closely match radiation emitted from the Sun. To the best of our knowledge, up to now, a working solar device in the infrared based on hot electron extraction has not been observed. However, such strategy is quite promising since it is using Earth's abundant and non-toxic materials. We also provided an overview of the tools which can be used to investigate the optical properties of those novel materials and devices. In the end, the harvesting of the UV region of the solar spectrum has been mentioned as a mean to slightly increase efficiency and maybe prolong the lifetime of devices suffering from UV degradation. Considering the constantly increasing demand of energy and the scarceness of resources on which global energy production is based on, any effort towards the fabrication of more efficient, greener devices would be of great importance.

# Acknowledgements

This project has received funding from the European Research Council (ERC) under the European Union's Horizon 2020 research and innovation programme (grant agreement No. [816313]).